\documentclass[conference]{IEEEtran}
\IEEEoverridecommandlockouts
\usepackage{cite}
\usepackage{amsmath,amssymb,amsfonts}
\usepackage{mathtools}
\usepackage{algorithmic}
\usepackage[linesnumbered, ruled]{algorithm2e}
\usepackage{graphicx}
\usepackage{textcomp}
\usepackage{xcolor}
\usepackage{xspace}
\usepackage{hyperref}
\usepackage{booktabs}
\usepackage{multirow}
\usepackage{subfigure}
\usepackage{makecell}
\usepackage{enumitem}
\usepackage[symbol]{footmisc}

\def\BibTeX{{\rm B\kern-.05em{\sc i\kern-.025em b}\kern-.08em
    T\kern-.1667em\lower.7ex\hbox{E}\kern-.125emX}}
\newcommand{\xhdr}[1]{{\bfseries #1}.}
\newcommand{\name}{BOURNE\xspace}
\newcommand{\jie}[1]{{{\textcolor{black}{#1}}}}

\begin{document}

\title{BOURNE: Bootstrapped Self-supervised Learning Framework for Unified Graph Anomaly Detection
}

\author{
\IEEEauthorblockN{Jie Liu}
\IEEEauthorblockA{School of Computer Science\\
    Northwestern Polytechnical University\\
    Xi'an, China\\
    jayliu@mail.nwpu.edu.cn}\\
\and
\IEEEauthorblockN{Mengting He}
\IEEEauthorblockA{School of Computer Science\\
    Northwestern Polytechnical University\\
    Xi'an, China\\
    hmt468@mail.nwpu.edu.cn}\\
\and
\IEEEauthorblockN{Xuequn Shang$^*$}
\IEEEauthorblockA{School of Computer Science\\
    Northwestern Polytechnical University\\
    Xi'an, China\\
    shang@nwpu.edu.cn}\\
\and
\IEEEauthorblockN{Jieming Shi}
\IEEEauthorblockA{Department of Computing\\
The Hong Kong Polytechnic University\\
    HongKong, China\\
    jieming.shi@polyu.edu.hk\\}
\and
\IEEEauthorblockN{Bin Cui}
\IEEEauthorblockA{School of Computer Science\\ 
    Peking University\\
    Beijing, China\\
    bin.cui@pku.edu.cn\\}
\and
\IEEEauthorblockN{Hongzhi Yin$^*$}
\IEEEauthorblockA{
School of Electrical Engineering \& Computer Science\\
The University of Queensland\\
    Brisbane, Australia\\
    h.yin1@uq.edu.au\\}
}

\maketitle

\begin{abstract}
Graph anomaly detection (GAD) has gained increasing attention in recent years due to its critical application in a wide range of domains, such as social networks, financial risk management, and traffic analysis. Existing GAD methods can be categorized into node and edge anomaly detection models based on the type of graph objects being detected. However, these methods typically treat node and edge anomalies as separate tasks, overlooking their associations and frequent co-occurrences in real-world graphs. As a result, they fail to leverage the complementary information provided by node and edge anomalies for mutual detection. Additionally, state-of-the-art GAD methods, such as CoLA and SL-GAD, heavily rely on negative pair sampling in contrastive learning, which incurs high computational costs, hindering their scalability to large graphs. To address these limitations, we propose a novel unified graph anomaly detection framework based on bootstrapped self-supervised learning (named \name). We extract a subgraph (graph view) centered on each target node as node context and transform it into a dual hypergraph (hypergraph view) as edge context. These views are encoded using graph and hypergraph neural networks to capture the representations of nodes, edges, and their associated contexts. By swapping the context embeddings between nodes and edges and measuring the agreement in the embedding space, we enable the mutual detection of node and edge anomalies.
Furthermore, BOURNE can eliminate the need for negative sampling, thereby enhancing its efficiency in handling large graphs. Extensive experiments conducted on six benchmark datasets demonstrate the superior effectiveness and efficiency of \name in detecting both node and edge anomalies.
\end{abstract}

\footnotetext[1]{Corresponding authors.}
%\begin{IEEEkeywords}
%component, formatting, style, styling, insert
%\end{IEEEkeywords}

\section{Introduction}
In recent years, graph anomaly detection (GAD) has received considerable attention due to its critical application in a wide range of domains. Typical application fields include social networks~\cite{ahmed2022combining, tam2019anomaly, sun2021heterogeneous}, traffic analysis~\cite{han2020traffic}, financial risk management~\cite{cheng2020graph}, and biological networks~\cite{mahmud2021deep}, to name a few. GAD seeks to identify anomalous graph objects, such as nodes, edges, or subgraphs, that significantly deviate from the regular patterns observed in most of the graph. In this work, we focus on detecting node and edge anomalies and divide the existing GAD approaches into node anomaly detection (NAD) and edge anomaly detection (EAD) methods. 

\begin{figure}[t]
    \centering
    \subfigure[Transaction network]{
    \includegraphics[scale=0.57]{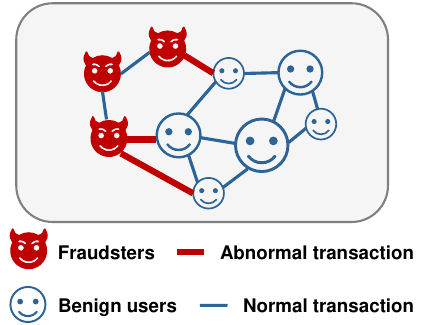}\label{fig:toy_a}
    }\hspace{-5mm}
    \subfigure[Dual hypergraph transformation]{
    \includegraphics[scale=0.57]{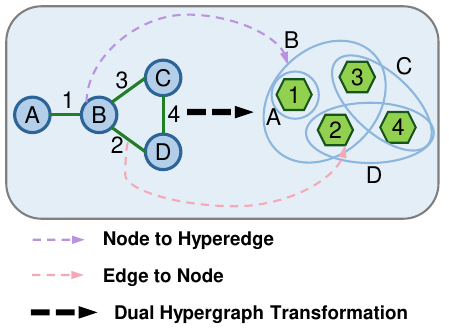}\label{fig:toy_b}
    }
    \caption{(a) A toy example of an online transaction network where abnormal transactions are commonly associated with fraudsters. (b) Illustration of dual hypergraph transformation. Here the numbers denote the edges in the graph and nodes in the hypergraph, while the letters denote the nodes in the graph and hyperedges in the hypergraph.}
    \label{fig:toy}
\end{figure}

Earlier anomalous node detection researches mainly resort to shallow learning mechanisms. For instance, Radar~\cite{li2017radar} utilizes residual analysis to detect anomalies by factorizing node attribute matrix. ANOMALOUS~\cite{peng2018anomalous} further employs CUR decomposition~\cite{mahoney2009cur} to select structure-related node attributes before residual analysis.
% For example, Radar~\cite{} utilizes residual analysis to detect anomalies by factorizing node attribute matrix. ANOMALOUS~\cite{} further employs CUR decomposition~\cite{} to select structure-related node attributes before computing residual errors. 
Despite their simplicity, shallow methods cannot capture the complex interactions inherent in graph data. In light of this, researchers have recently introduced deep learning models into the anomalous node detection task. DOMINANT~\cite{ding2019deep} employs a graph autoencoder to reconstruct both the structural and attributive information of graphs and then evaluate the node abnormality by reconstruction error. CoLA~\cite{liu2021anomaly} constructs instance pairs consisting of nodes and subgraphs and applies contrastive learning to learn node representation. The anomaly score of a node is then calculated by measuring the agreement between the pair of instances. SL-GAD~\cite{zheng2021generative} further integrates both generative attribute reconstruction and contrastive learning. 

As for anomalous edge detection, they usually first learn node-level representations and then measure the edge abnormality with the embeddings of the nodes it connects. For example, AANE~\cite{duan2020aane} first generates node embeddings using GCN~\cite{kipf2016semi} and then calculates the link probability as the hyperbolic tangent of the connected node embeddings. An edge is deemed anomalous if its predicted probability falls below a predefined threshold. UGED~\cite{ouyang2020unified} utilizes an autoencoder and a fully connected network to predict edge distribution and identify edge abnormality through its appearance probability.

% In contrast, \name takes a unified approach by leveraging both node and edge anomaly information simultaneously. In view of the common co-existence of node and edge anomalies, this joint consideration can enhance the mutual detection of each other.

Despite the success of existing GAD methods, they still suffer from two major limitations:
(1) Existing GAD methods typically approach anomalous node and edge detection as independent tasks, failing to consider their associations in real-world datasets (\textbf{P1}). For instance, as shown in Figure \ref{fig:toy_a}, in online transaction networks, abnormal transactions (anomalous edges) are often associated with fraudsters (anomalous nodes)~\cite{cheng2020graph, zhang2020gcn}, while in social networks, social spammers (anomalous nodes) frequently send spam messages (anomalous edges) to regular users~\cite{deng2022markov}. Due to the strong correlation between two types of anomalies, detecting anomalous patterns of nodes can provide valuable auxiliary information for edge anomaly detection, and vice versa. However, none of the existing methods can detect both types of anomalies simultaneously, and utilize their mutual detection capabilities. (2) The current state-of-the-art GAD models, such as CoLA~\cite{liu2021anomaly} and SL-GAD~\cite{zheng2021generative}, employ contrastive learning techniques that aim to maximize agreement among elements within the same instance pairs while minimizing agreement among elements from different pairs. Thus, they heavily rely on extensive negative instance pairs to detect anomalies. Consequently, generating and processing a large number of negative pairs brings extra computational cost, which becomes prohibitively high when dealing with large-scale graphs (\textbf{P2}).
In order to address the limitations above, we propose a novel \underline{\textbf{BO}}otstrapped Self-supervised Learning Framework for \underline{\textbf{U}}nified G\underline{\textbf{R}}aph A\underline{\textbf{N}}omaly D\underline{\textbf{E}}tection (\textbf{\name} for abbreviation) that efficiently detects anomalous nodes and edges simultaneously.
The key idea behind \name is to construct discrimination pairs for both target nodes and target edges, employ self-supervised learning to compare these pairs, and derive anomaly scores. To achieve this, we initially extract enclosing subgraphs centered on each target node as a graph view (node context). Then, we apply dual hypergraph transformations with augmentations to the graph view, resulting in a hypergraph view (edge context). As illustrated in Figure \ref{fig:toy_b}, an edge in the original graph view corresponds to a node in the hypergraph view, and a node in the graph view corresponds to a hyperedge in the hypergraph view.
Both views are then passed through a Graph Neural Network (GNN) encoder and a Hypergraph Neural Network (HGNN) encoder, respectively, to learn latent representations for nodes and edges. We facilitate mutual detection by swapping the surrounding contexts of target nodes and target edges. Specifically, we compute the anomaly score of a target node by comparing its representation from a graph view with the context representations from a hypergraph view. Similarly, we calculate the anomaly score for a target edge by comparing edge representation from the hypergraph view with the context representations from the graph view.
Finally, by combining both node and edge anomaly scores, we obtain a comprehensive objective. \jie{\cite{grill2020bootstrap} and \cite{thakoor2021large} initially introduced bootstrapped self-supervised learning frameworks to obviate the necessity of negative sampling, but tailored for image and node classification domains, making them not directly applicable to anomaly detection. We present a specialized bootstrapped self-supervised model tailored for graph anomaly detection, demonstrating strong performance without the need of negative pair sampling. Thereby, the proposed model can efficiently handle large-scale graphs.}

In summary, the main contributions of our work are as follows:
\begin{itemize}
    \item \jie{We propose a unified graph anomaly detection model}, \name, that leverages the correlation of anomalous nodes and edges to enhance the detection of each other. To the best of our knowledge, this is the first unified framework capable of detecting both types of anomalies simultaneously.
    % \item We propose a straightforward yet powerful metric called \textit{anomaly correlation} that accurately measures the correlations between anomalous nodes and anomalous edges. This metric serves to justify the applicability of our model in detecting and understanding the relationships between these anomalous objects.
    \item \jie{We develop a bootstrapped self-supervised learning framework for the graph anomaly detection task, and it significantly improves the learning efficiency without the need for negative pair sampling.} The proposed model can efficiently handle large-scale datasets without sacrificing performance.
    \item Extensive experiments on six benchmark datasets show that \name achieves superior performance on both node and edge anomaly detection compared with state-of-the-art (SOTA) baselines.
\end{itemize}

% Despite their success, there are two main limitations of the existing GAD methods. First, anomalous nodes and edges often coexist in real-world datasets. For example, fraudulent transactions (abnormal edges) mostly happen between fraudsters and normal users in online transaction networks, as shown in Figure~\ref{}(a). However, none of the current methods have considered the potential benefit of leveraging the co-occurrence of anomalous nodes and edges to enhance the mutual detection of each other.

\section{Related Work}
In this section, we introduce the works closely related to ours: Graph and Hypergraph Neural Networks, Self-supervised Learning, and Graph Anomaly Detection.

\subsection{Graph and Hypergraph Neural Networks}
Network embedding aims to project the nodes or edges into dense latent vectors while considering both the structural and attributive information of the nodes or edges\cite{liu2023imbalanced}. Graph neural networks (GNNs) have emerged as one of the most powerful network embedding approaches due to their ability to conduct deep learning on non-Euclidean graph data~\cite{chen2019exploiting}. Spectral CNN~\cite{bruna2013spectral} first defines the spectral convolution operation on the graph domain by employing filters on graph signals. Following~\cite{bruna2013spectral}, ChebNet~\cite{defferrard2016convolutional} performs localized approximation of the filter by Chebyshev polynomials of the eigenvalues. GCN~\cite{kipf2016semi} further introduces a first-order approximation of ChebNet. GraphSage~\cite{hamilton2017inductive} defines a message passing process from neighbors to update the central node's representation and proposes mini-batch training for model scalability. GAT~\cite{velivckovic2017graph} introduces an attention mechanism to aggregate neighbor messages with adaptive weights.

A hypergraph is a generalization of a graph in which a hyperedge can join an arbitrary number of vertices, so that can model high-order interactions. Similar to GNNs, recent research has also extended deep learning to hypergraphs, giving rise to hypergraph neural networks (HGNNs). HGNN~\cite{feng2019hypergraph} first defines convolution on hypergraphs with hypergraph Laplacian matrix~\cite{zhou2006learning}. HyperGCN~\cite{yadati2019hypergcn} simplifies the scheme by approximating each hyperedge with a pairwise edge. \cite{sun2021heterogeneous} further introduces wavelet basis to hypergraph convolution.

% In our proposed \name, we select GCN~\cite{kipf2016semi} and HGNN~\cite{feng2019hypergraph} as the backbone to encode the information of constructed graph view and hypergraph view, respectively. They can be flexibly replaced with more sophisticated graph or hypergraph neural networks.

\subsection{Self-supervised Learning}
In real-world situations, the scarcity or imbalance of accurate labels often poses a challenge~\cite{hao2021pre}. Self-supervised learning (SSL) addresses this by allowing the training of deep models on unlabeled data~\cite{yu2023self}. As a branch of SSL, contrastive learning has made significant performance strides and recent research has extended contrastive frameworks to accommodate graph data. These approaches begin by creating instance pairs from the input graph and subsequently train the encoders to ensure that the representations of positive instance pairs are in agreement, while the representations of negative instance pairs exhibit disagreement. DGI~\cite{velivckovic2018deep} employs graph corruption to generate negative samples and utilizes node-level and graph-level representations as instance pair for contrast. MVGRL~\cite{hassani2020contrastive} and GCC~\cite{qiu2020gcc} contrast views between nodes and either subgraphs or structurally transformed graphs and utilizes the Jensen-Shannon estimator for learning. 
Apart from general node representation, contrastive learning has also been applied in GAD recently by extracting a node and its surrounding subgraph as an instance pair and assessing the disagreement between them as an anomaly score~\cite{liu2021anomaly, zheng2021generative}.

While these contrastive learning models have achieved remarkable success, their performance heavily relies on comparing with a significant number of negative pairs. This requirement can lead to computational and memory burdens that scale quadratically with the number of nodes. Consequently, it hinders the scalability of these models to deal with large graphs.

\subsection{Graph Anomaly Detection}
Graph anomaly detection (GAD) aims at identifying anomalous graph objects (i.e., nodes, edges, and subgraphs) in the graph~\cite{ma2021comprehensive, wang2021decoupling}. 
% Based on the types of anomalous objects, GAD can be categorized into anomalous node detection, anomalous edge detection, and anomalous subgraph detection. 
Due to the varying sizes and intricate internal structures of anomalous subgraphs~\cite{sun2023tinyad}, we leave this challenging problem for future research and mainly focus on anomalous node and edge detection in this work. Early anomalous node detection approaches mainly use shallow techniques such as residual analysis (Radar~\cite{li2017radar}), matrix factorization (ALAD~\cite{liu2017accelerated}), and CUR decomposition (ANOMALOUS~\cite{peng2018anomalous}) to extract anomalous pattern in graphs. More recently, \cite{ding2019deep} pioneered the integration of deep learning into node anomaly detection by employing a graph autoencoder to reconstruct both the structure and attribute information of graphs. The anomaly score for a node is then determined based on the reconstruction errors. CoLA~\cite{liu2021anomaly} and SL-GAD~\cite{zheng2021generative} further introduce graph contrastive learning that captures abnormal patterns by learning agreements between nodes and subgraphs. As for anomalous edge detection, UGED~\cite{ouyang2020unified} proposes to model the distribution of edges through an autoencoder and fully connected network and identify the edges that are least likely to appear as anomalies. AANE~\cite{duan2020aane} first generates node embeddings using GCN, and then calculates the link probability as the hyperbolic tangent of the connected node embeddings. An edge is considered anomalous if its predicted probability is below a predefined threshold.

Apparently, existing GAD methods focus only on individual abnormal nodes or edges. There is a lack of methods that can simultaneously detect both types of anomalies with acceptable resources. Moreover, none of the methods have considered the potential benefit of leveraging the co-occurrence of both types of anomalies to enhance their mutual detection.

\begin{table}[t!]
  \setlength\tabcolsep{2.0pt}
  \centering
  \caption{Summary of Notations.}\label{tab:nota}
  \begin{tabular}{l|l}
    \toprule
    \textbf{Notations} & \textbf{Descriptions} \\
    \midrule
    $\mathcal{G}=\{\mathbf{X}, \mathbf{A}\}$ & An attributted graph. \\
    $\mathcal{G}^*=\{\mathbf{X}^*, \mathbf{M}^T\}$ & A dual hypergraph of $\mathcal{G}$. \\
    $\mathcal{V}, \mathcal{E}$ & The node and edge set of $\mathcal{G}$. \\
    $\mathbf{A}\in\mathbb{R}^{N\times N}$ & The adjacency matrix of $\mathcal{G}$. \\
    $\mathbf{M}\in\mathbb{R}^{N\times M}$ & The incidence matrix of $\mathcal{G}$. \\
    $\mathbf{X}\in\mathbb{R}^{N\times D}$ & The node feature matrix of $\mathcal{G}$. \\
    $\mathbf{y}_n\in\mathbb{R}^N, \mathbf{y}_e\in\mathbb{R}^M$ & Anomalous label vector of nodes and edges. \\
    $\mathbf{M}^*=\mathbf{M}^\mathrm{T}\in\mathbb{R}^{M\times N}$ & The incidence matrix of $\mathcal{G}^*$. \\
    $\mathbf{X}^*\in\mathbb{R}^{M\times D}$ & The node feature matrix of $\mathcal{G}^*$. \\
    \midrule
    $v_t$, $\mathcal{E}_t$ & Target node and target edges. \\
    $\mathcal{N}(\cdot)$ & 1-hop neighbors of a node. \\
    $\mathcal{G}_t, \mathcal{G}^*_t$ & Graph and hypergraph view of $v_t$ and $\mathcal{E}_t$. \\
    $\mathbf{H}^{(l)}\in \mathbb{R}^{(N_{s}+1)\times D'}$ & The node embedding matrix of $\mathcal{G}_t$. \\
    $\mathbf{Z}^{(l)}\in \mathbb{R}^{(M_{s}+M_{tar})\times D'}$ & The edge embedding matrix of $\mathcal{G}^*_t$. \\
    $\mathbf{Z}_t$ & The embedding matrix of target edges. \\
    $\mathbf{Z}_p, \mathbf{Z}_s\in\mathbb{R}^{D'}$ & Patch- and subgraph-level embedding matrix of $\mathcal{E}_t$.\\
    $\bar{\mathbf{h}}_t$ & The embedding of target node. \\
    $\bar{\mathbf{h}}_p, \bar{\mathbf{h}}_s\in\mathbb{R}^{D'}$ & Patch- and subgraph-level embedding vector of $v_t$.\\
    \midrule
    $N$ & The number of nodes in $\mathcal{G}$. \\
    $M$ & The number of edges in $\mathcal{G}$. \\
    $N_s$ & The number of nodes in $\mathcal{G}_t$. \\
    $M_s$ & The number of edges in $\mathcal{G}_t$. \\
    $M_{tar}$ & The number of target edges in $\mathcal{G}_t$. \\
    $D$ & The dimension of node/edge features in $\mathcal{G}$. \\
    $D'$ & The dimension of latent node/edge embeddings. \\
    
    \bottomrule
  \end{tabular}
\end{table}

\section{Preliminaries}
In this section, we provide the definitions of essential concepts and formalize the problem of self-supervised graph anomaly detection. We also summarize the frequently used notations across this paper in Table \ref{tab:nota} for quick reference.
% Notation Table here

\xhdr{Definition 1: Attributed Graph}
Given an attributed graph $\mathcal{G}=\{\mathbf{X}, \mathbf{A}\}$, we use $\mathbf{X}\in\mathbb{R}^{N\times D}$ to denote the node feature matrix and its $i$-th row vector $\mathbf{x}_i\in\mathbb{R}^D$ represents the feature of node $v_i$. $\mathcal{V}=\{v_1, v_2,\ldots,v_N\}$ and $\mathcal{E}=\{e_1, e_2,\ldots,e_M\}$ are node and edge set of the graph, respectively. The structural information of $\mathcal{G}$ is represented by the graph adjacency matrix $\mathbf{A}\in\mathbb{R}^{N\times N}$, where $\mathbf{A}_{ij}=1$ if $e_{ij}=(v_i,v_j)\in\mathcal{E}$, otherwise $\mathbf{A}_{ij}=0$. There are also two binary label vectors $\mathbf{y}_n\in\mathbb{R}^N$ and $\mathbf{y}_e\in\mathbb{R}^M$ indicating the anomalous state of nodes and edges, respectively. In these vectors, the value of 1 represents an abnormal state, while 0 denotes a normal state.

\xhdr{Definition 2: Dual Hypergraph Transformation}\label{def:dual}
Given an attributed graph $\mathcal{G}=\{\mathbf{X}, \mathbf{A}\}$, we can also represent its structural information as incidence matrix $\mathbf{M}\in \mathbb{R}^{N\times M}$, which represents the interaction between $N$ nodes and $M$ edges. Each entry in $\mathbf{M}$ indicates whether the node is incident to the edge. By transforming the edges and nodes of the original graph $\mathcal{G}=\{\mathbf{X}, \mathbf{M}\}$ into the nodes and hyperedges of the dual hypergraph, we obtain the dual hypergraph $\mathcal{G}^*$ of $\mathcal{G}$, which consists of $M$ nodes and $N$ hyperedges. $\mathcal{G}^*$ can be formally defined as $\mathcal{G}^*=\{\mathbf{X}^*, \mathbf{M}^*\}$, where $\mathbf{X}^*[t,:]=\frac{1}{2}(\mathbf{X}[i,:]+\mathbf{X}[j,:]), e_t=(v_i, v_j)\in\mathcal{E}$, and $\mathbf{M}^*=\mathbf{M}^T$.

\xhdr{Problem Formulation: Self-supervised Graph Anomaly Detection}
Given an attributed graph $\mathcal{G}=\{\mathbf{X},\mathbf{A}\}$, our goal is to learn a comprehensive anomaly detection model $\mathcal{F}=\{\mathcal{F}_n, \mathcal{F}_e\}$, where $\mathcal{F}_{n}(\cdot):\mathbb{R}^{N\times D}\rightarrow\mathbb{R}^{N\times 1}$, and $\mathcal{F}_e(\cdot):\mathbb{R}^{M\times D}\rightarrow\mathbb{R}^{M\times 1}$. It can simultaneously measure the degree of abnormality of both nodes and edges in $\mathcal{G}$ by calculating their anomaly scores in a unified way. Since we consider self-supervised GAD in this paper, none of the node or edge labels indicating abnormality is available during the training phase. 

% \xhdr{Definition 2: Anomaly Correlation} Given an attributed graph $\mathcal{G}$ with anomalous labels $\mathbf{Y}_n$ and $\mathbf{Y}_e$. The anomaly correlation $C_{ano}$ between anomalous nodes and edges can be defined as the conditional probability $P(e_a\mid v_a)$:
% \begin{equation}
%     P(e_a\mid v_a)=\frac{P(e_a,v_a)}{P(v_a)},
% \end{equation}
% where $P(e_a,v_a)$ and $P(v_a)$ can be formed as follows:
% \begin{align}
%     P(v_a)&=\frac{\left|\mathcal{V}_a\right|}{\left|\mathcal{V}\right|},\label{eq:yn}\\
%     P(e_a,v_a)&=\frac{1}{\left|\mathcal{V}\right|}\sum_{v\in\mathcal{V}}\frac{\left|e\in N(v):y_e=y_v=1\right|}{D(v)},\label{eq:ynye}
% \end{align}
% where $\mathcal{V}_a$ is the anomalous node set. Combining equations \ref{eq:yn} and \ref{eq:ynye}, $C_{ano}\in[0,1]$ can be formed as:
% \begin{equation}
%     C_{ano}=P(e_a\mid v_a)=\frac{1}{\left|\mathcal{V}_a\right|}\sum_{v\in\mathcal{V}_a}\frac{\left|e\in N(v):y_e=y_v=1\right|}{D(v)}.
% \end{equation}
% Graphs exhibiting strong correlation between node and edge anomalies have higher value of $C_{ano}$, and vice versa.

\section{Methods}
\begin{figure*}[t!]
    \centering
    \includegraphics[scale=0.73]{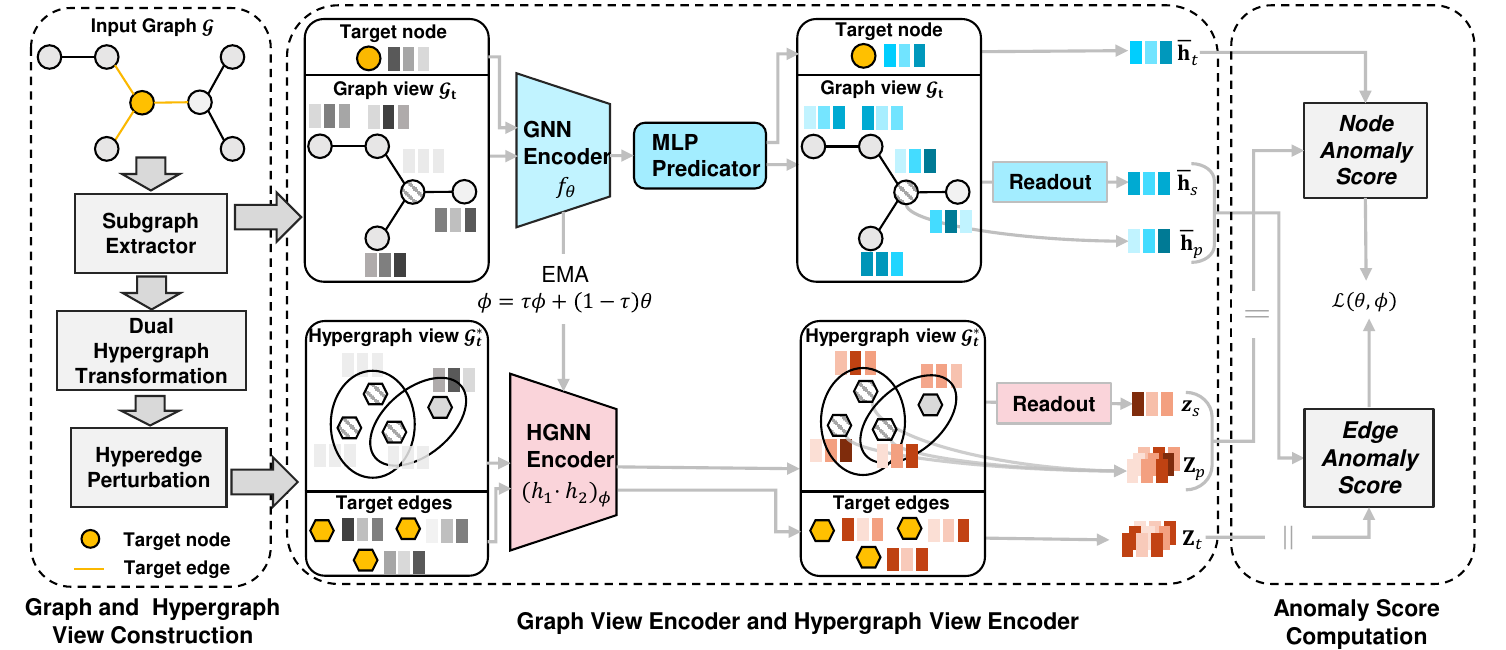}
    \caption{Overall framework of the proposed \name.}
    \label{fig:frame}
\end{figure*}

In this section, we introduce our proposed \name which can detect both node-level and edge-level anomalies in an unsupervised manner. As illustrated in Figure \ref{fig:frame}, \name consists of three components: (1) \textit{Graph and Hypergraph View Construction} that constructs a graph view for node anomaly detection and a hypergraph view for edge anomaly detection, respectively. (2) \textit{Unified graph view encoder} which consists of two encoding channels, i.e., \textit{graph view encoder} to encode target nodes and their surrounding contexts in the graph view into embeddings and \textit{hypergraph view encoder} to encode the target edges and their contexts in the hypergraph view into embeddings. (3) \textit{Graph Anomaly Score Computation} that measures the abnormality of a target node or edge by calculating both the patch-level and subgraph-level agreement in the embedding space. 
% In this way, if a selected target node/edge is an anomaly, the attributive and structural mismatch between it and its contexts can be reflected by the patch- and subgraph-level disagreement, and thus be assigned with higher anomaly scores.

% Specifically, for each node from the input graph, \textit{Graph and Dual Hypergraph View Construction} first extracts a local subgraph for it and then augments the subgraph into a graph view for node anomaly detection and a hypergraph view for edge anomaly detection, respectively. After that, to extract the node and edge anomaly signals to supervise each other and conduct self-supervised anomaly detection, we construct a \textit{unified graph view encoder} which consists of two encoding channels, i.e., \textit{graph view encoder} to encode graph view information to acquire low-dimensional embeddings of the target node and its surrounding contexts,  and \textit{hypergraph view encoder} to encode the information from hypergraph view to obtain embeddings of target edges and its contexts. After that, the final component \textit{node and edge anomaly score computation} measures the abnormality of a target node or edge by calculating both the patch-level (i.e., node v.s. node or edge v.s. edge) and subgraph-level (i.e., node v.s. subgraph or edge v.s. subgraph) agreement in the embedding space. In this way, if a selected target node/edge is an anomaly, the attributive and structural mismatch between it and its contexts can be reflected by the patch- and subgraph-level disagreement, and thus be assigned with higher anomaly scores.

In the rest of this section, we introduce the three components in detail from section \ref{sub:view} to \ref{sub:score}. Then we analyze the complexity of our model in subsection \ref{sub:comp}. The overall pipeline of \name is illustrated in Algorithm \ref{alg:1}.

\subsection{Graph and Dual Hypergraph View Construction}\label{sub:view}
%demonstrate the principle of current GAD models.
% Self-supervised graph representation learning models usually define various types of instance pairs to provide rich structural and attributive information for self-supervised learning. By encoding the discrimination pairs into the embedding space, the abnormality of a target structure can be reflected by the dissimilarity degree of the obtained pair of embeddings.

In the context of graph anomaly detection, the graph object (i.e., node or edge) that we aim to evaluate for its abnormality is referred to as a ``target object". 
% Since the self-supervised learning setting lacks explicit labels, we rely on the intrinsic attributes and structural characteristics of the graph to detect anomalies.
In our approach, we distinguish anomalous objects from the majority of normal ones by examining the inconsistency between the target object and its surrounding contexts. These contexts, along with the target object itself, form \textit{discrimination pairs}. Depending on the size of the constructed context, the discrimination can be performed either between instances of the same scale or across different scales. To comprehensively assess the discrimination of various scales, we establish discrimination pairs at both the patch-level (i.e., node v.s. node or edge v.s. edge) and subgraph-level (i.e., node v.s. subgraph or edge v.s. subgraph).

To achieve this, we first extract an enclosing subgraph for every target node because the local subgraph can provide adequate information on both patch-level and subgraph-level contexts for robust statistical anomaly scoring. Then, instead of indirectly obtaining edge embeddings through node embeddings as done in previous works~\cite{duan2020aane,ouyang2020unified}, we transform the extracted subgraph into a dual hypergraph. After applying simple augmentations to the dual hypergraph, the augmented hypergraph, together with the extracted subgraph, serve as two views for training our self-supervised GAD model.

In the following, we further elaborate on the pre-processing steps mentioned above.

\xhdr{Targe node and target edge sampling} Since we aim at detecting both anomalous nodes and edges in a unified way, target nodes and edges need to be sampled first. Specifically, we randomly sample a target node $v_t$ from the node set $\mathcal{V}$ of the input graph $\mathcal{G}$ through uniform sampling without replacement. Then all the edges connected to the target node $e\in\mathcal{E}_t=\{(v_i, v_t)\mid v_i \in \mathcal{N}(v_t)\}$ are selected as target edges. Here $\mathcal{N}(v_t)$ indicates the 1-hop neighbor set of $v_t$.

\xhdr{Subgraph extraction} We then extract the enclosing subgraph $\mathcal{G}_t$ for each $v_t$. Previous works~\cite{liu2021anomaly, zheng2021generative} set the target node as the initial node and apply random walk with restart (RWR) to construct the subgraph. However, this procedure may exclude too many target edges from the subgraph and cannot fulfill the edge anomaly detection task. Therefore, we directly sample $K$ nodes from the $k$-hop neighbors of $v_t$ with replacement. This approach ensures that all the enclosing subgraphs have a fixed size of $K+1$ nodes while preserving as many target edges as possible. We directly use the subgraph $\mathcal{G}_t$ without augmentation as the graph view.

\xhdr{Dual hypergraph transformation} Since accurately learning the edge-level representations is crucial to the success of edge anomaly detection, here we explicitly conduct message-passing for edges to learn optimal representations of edges. To achieve this, inspired by the hypergraph duality~\cite{scheinerman2011fractional} and edge representation with hypergraph~\cite{jo2021edge}, we transform the edges and nodes of the original subgraph $\mathcal{G}_t$ into the nodes and hyperedges of a hypergraph. The dual hypergraph transformation procedure is elaborated in Preliminaries \ref{def:dual}. After that, we obtain the dual hypergraph $\mathcal{G}^*_t$.

\xhdr{Hypergraph perturbation} \jie{The selection of augmentation strategies can highly affect the performance of self-supervised models. Previous bootstrapped self-supervised models~\cite{thakoor2021large} use a combination of node feature masking and edge masking as the augmentation function. However, this strategy may omit target edges from the subgraph, introducing extra noises and hindering anomaly detection.}
% Since the structural and attributive information of the extracted subgraph is equivalent to the transformed dual hypergraph~\cite{wei2022augmentations},
To address the problem, we fabricate two simple augmentations $\mit{\Gamma}_1$ and $\mit\Gamma_2$ on the dual hypergraph. Specifically, $\mit\Gamma_1$ is node feature masking, which stochastically masks the input node features of the hypergraph as zero. $\mit\Gamma_2$ is hyperedge perturbation. To avoid introducing extra anomalous nodes or edges, here we only randomly kick out nodes from the hyperedges following an i.i.d. Bernoulli distribution based on the hypothesis that partial disruption of high-order relations does not significantly affect the semantics of hyperedge representations~\cite{wei2022augmentations}. \jie{In this way, the node number within the hypergraph stays constant after augmentation.} The hypergraph view is then obtained by $\mathcal{G}^*_t=\mit{\Gamma}_2(\mit{\Gamma}_1(\mathcal{G}^*_t))$.

% \textbf{Anonymization} To prevent the information leakage of the target node/edge and encourage the model to detect anomalies by relying on contextual information, we anonymize the target nodes and edges by masking their initial features to zero vectors.

\subsection{Graph View Encoder}\label{sub:uni}
% In this section, we aim at learning the embeddings of the target node and surrounding contexts by encoding the constructed graph view $\mathcal{G}_k=\{\mathcal{V}_k, \mathcal{E}_k, \mathbf{X}\}$. 
As stated in Section \ref{sub:view}, we conduct a comprehensive anomaly scoring by comparing the dissimilarity of the target node with both patch-level and subgraph-level contexts. The patch-level discrimination involves calculating the agreement between $\mathbf{h}_{p}$, the embedding of the target node $v_t$ after aggregating the features of its neighboring nodes, and $\mathbf{h}_t$, the embedding of $v_t$ with independent projection. On the other hand, the subgraph-level discrimination aims to learn the agreement between $\mathbf{h}_s$, the embedding of the subgraph $\mathcal{G}_t$ obtained through the readout module, and the independent embedding $\mathbf{h}_t$. However, the attribute of $v_t$ would be involved in the calculation of $\mathbf{h}_p$ and $\mathbf{h}_s$, which can potentially lead to information leakage and increase the difficulty of detecting abnormal nodes through contextual discrimination. To mitigate this issue, we first anonymize $v_t$ in the graph view and subsequently perform graph view encoding.

\xhdr{Target Node Anonymization} Given the target node $v_t$ and the extracted surrounding subgraph $\mathcal{G}_t=\{\mathbf{X}_t, \mathbf{A}_t\}$, we first anonymize $v_t$ by masking its feature as zero vector and then concatenate its feature in the end. Specifically, the anonymized feature matrix $\widehat{\mathbf{X}}_t\in \mathbb{R}^{(N_{s} + 1)\times D}$ is obtained as follows:
\begin{equation}
    \widehat{\mathbf{X}}_t = \textit{CONCAT}\bigl(\vec{\boldsymbol{0}}_{1\times D}, \mathbf{X}_t[2:N_{s},:], \mathbf{X}_t[1,:]\bigr)
\end{equation}
where $\vec{\boldsymbol{0}}_{1\times D}\in \mathbb{R}^{1\times D}$ denotes the zero vector and $N_{s}$ is the node number of the subgraph. To ensure the independence of $\mathbf{h}_t$ without interference from neighboring information, we introduce an isolated entry to the adjacency matrix $\mathbf{A}_t\in\mathbb{R}^{N_{s}\times N_{s}}$ and acquire $\widehat{\mathbf{A}}_t\in\mathbb{R}^{(N_{s}+1)\times (N_{s}+1)}$ in below:
\begin{equation}
\widehat{\mathbf{A}}_t = \begin{bmatrix*}[l]
\mathbf{A}_t & \boldsymbol{0}_{N_{s}\times1} \\
\boldsymbol{0}_{1\times N_{s}} & 1 \\
\end{bmatrix*}
\end{equation}

Thus we obtain the anonymized graph view $\widehat{\mathcal{G}}_t=\{\widehat{\mathbf{X}}_t,\widehat{\mathbf{A}}_t\}$.

\xhdr{Graph Encoder} The graph encoder takes the anonymized subgraph $\widehat{\mathcal{G}}_t$ as input, and outputs the patch-level context embedding $\bar{\mathbf{h}}_p$, subgraph-level context embedding $\bar{\mathbf{h}}_s$ and target node embedding $\bar{\mathbf{h}}_t$. Specifically, the representation of nodes in $\widehat{\mathcal{G}}_t$ is computed as follows:
\begin{equation}\label{eq:gnn}
    \mathbf{H}^{(l)}=\textit{GNN}_{\theta}\Bigl(\widehat{\mathbf{A}}_t, \mathbf{H}^{(l-1)}; \mathbf{\Theta}^{(l)}\Bigr),
\end{equation}
where $\mathbf{\Theta}^{(l)}\in\mathbb{R}^{D\times D'}$ denotes the learnable parameters of the $l$-th layer. $\mathbf{H}^{(l-1)}$ and $\mathbf{H}^{(l)}$ are the node representation matrices learned by the $(l-1)$-th and $(l)$-th layer, respectively. With an $L$-th layer GNNs, the output node embedding is denoted as $\mathbf{H}^{(L)}$ and the input representation $\mathbf{H}^{(0)}$ is defined as the anonymized feature matrix $\widehat{\mathbf{X}}_t$. $\textit{GNN}_{\theta}(\cdot)$ can be set as any off-the-shelf graph neural networks. For computation efficiency, we adopt a one-layer graph convolutional network (GCN) as the backbone. Thus, Equation (\ref{eq:gnn}) can be specifically re-written as follows:
\begin{equation}\label{eq:gcn}
    \mathbf{H}^{(l)} = \sigma\Bigl(\widetilde{\mathbf{D}}^{-\frac{1}{2}}_t\widetilde{\mathbf{A}}_t\widetilde{\mathbf{D}}^{-\frac{1}{2}}_t\mathbf{H}^{(l-1)}\mathbf{\Theta}^{(l)}\Bigr),
\end{equation}
where $\widetilde{\mathbf{A}}_t=\widehat{\mathbf{A}}_t+\mathbf{I}_{N_{s}}$, and $\widetilde{\mathbf{D}}_t$ is a diagonal node degree matrix where $\widetilde{\mathbf{D}}_t(i,i)=\sum_j\widetilde{\mathbf{A}}_t(i,j)$. $\sigma(\cdot)$ is the activation function and \textit{PReLU} is adopted here. The $L$-th layer representation $\mathbf{H}^{(L)}$ is then fed into a predictor $p_{\theta}$ that outputs the prediction of the node representation $\overline{\mathbf{H}}$:
\begin{equation}
    \overline{\mathbf{H}}=p_\theta\bigl(\mathbf{H}^{(L)}\bigr). \label{eq:g_emb_1}
\end{equation}

Here we choose a two-layer multilayer perceptron (MLP) as $p_\theta$. The patch-level context embedding $\bar{\mathbf{h}}_p$ and the target node embedding $\bar{\mathbf{h}}_t$ can be easily obtained via $\bar{\mathbf{h}}_p = \overline{\mathbf{H}}[1,:]$ and $\bar{\mathbf{h}}_t = \overline{\mathbf{H}}[N_{s}+1,:]$, respectively. The subgraph-level context embedding $\bar{\mathbf{h}}_s$ is obtained with an average readout function:
\begin{equation}
    \bar{\mathbf{h}}_s=\textit{Readout}\Bigl(\overline{\mathbf{H}}\Bigr)=\frac{1}{N_{s}}\sum^{N_{s}}_{i=1}\overline{\mathbf{H}}[i,:]. \label{eq:g_emb_2}
\end{equation}

\subsection{Hypergraph View Encoder}
Through dual hypergraph transformation, nodes and hyperedges in the dual hypergraph correspond to edges and nodes in the original subgraph. Thus, we can apply any off-the-shelf message-passing schemes designed for node representation learning, for learning edge representations. To avoid confusion, in the subsequent sections of the paper, we consistently use ``edge'' to refer to the node in the hypergraph. Similar to graph view encoder, we also first anonymize the target edges $e_t\in \mathcal{E}_{tar}$ and then perform hypergraph view encoding to obtain target edge embeddings $\mathbf{Z}_t$, edge-level context embeddings $\mathbf{Z}_p$ and hypergraph-level context embeddings $\mathbf{Z}_s$, respectively.

\xhdr{Target Edge Anonymization} Given the target edge set $\mathcal{E}_{tar}$ and the dual hypergraph $\mathcal{G}^*_t=\{\mathbf{X}^*_t,\mathbf{M}^*_t\}$, the edge feature matrix $\mathbf{X}^*_t$ is anonymized by zero-masking the target edge features and concatenating them in the end, formally as:
\begin{equation}\footnotesize
    \widehat{\mathbf{X}}^*_t=\textit{CONCAT}\Bigl(\vec{\boldsymbol{0}}_{M_{tar}\times D}, \mathbf{X}^*_t[M_{tar}+1:M_s,:],\mathbf{X}^*_t[1:M_{tar},:]\Bigr),
\end{equation}
where $\widehat{\mathbf{X}}^*_t\in\mathbb{R}^{(M_s+M_{tar})\times D}$ denotes the anonymized edge feature matrix, $M_s$ and $M_{tar}$ are the number of edges and target edges of the hypergraph, respectively. $\vec{\boldsymbol{0}}_{M_{tar}\times d}\in\mathbb{R}^{M_{tar}\times D}$ is zero matrix. To encode the independent embeddings of target edges $\mathbf{Z}_t$, we also introduce isolated entries to the incidence matrix $\mathbf{M}^*_t\in\mathbb{R}^{M_s\times N_s}$ and acquire $\widehat{\mathbf{M}}^*_t\in\mathbb{R}^{(M_s+M_{tar})\times (N_s+M_{tar})}$ as follows:
\begin{equation}
    \widehat{\mathbf{M}}^*_t = \begin{bmatrix*}[l]
\mathbf{M}^*_t & \mathbf{0}_{M_{s}\times M_{tar}} \\
\mathbf{0}_{M_{tar}\times N_{s}} & \mathbf{I}_{M_{tar}} \\
\end{bmatrix*},
\end{equation}
where $\mathbf{I}_{M_{tar}}\in\mathbb{R}^{M_{tar}\times M_{tar}}$ denotes the identity matrix. The anonymized hypergraph view is denoted as $\widehat{\mathcal{G}}^*_t=\{\widehat{\mathbf{X}}^*_t,\widehat{\mathbf{M}}^*_t\}$.

\xhdr{Hypergraph Encoder} The hypergraph encoder then takes the anonymized hypergraph view $\widehat{\mathcal{G}^*_t}$ as input, and generate target edge embeddings $\mathbf{Z}_t$, edge-level context embeddings $\mathbf{Z}_p$ and hypergraph-level context embedding $\mathbf{z}_s$, respectively. We define the hypergraph encoder as follows:
\begin{align}
    \mathbf{Z}^{(l)}&=\textit{HGNN}_\phi\Bigl(\widehat{\mathbf{M}}^*_t, \mathbf{Z}^{(l-1)};\mathbf{\Phi}^{(l)}\Bigr)\\
    &=\sigma\Bigl(\mathbf{D}^{-\frac{1}{2}}_v\widehat{\mathbf{M}}^*_t\mathbf{W}_e\mathbf{D}^{-1}_e\widehat{\mathbf{M}}^{*T}_t\mathbf{D}^{-\frac{1}{2}}_v\mathbf{Z}^{(l-1)}\mathbf{\Phi}^{(l)}\Bigr),\label{eq:hgnn}
\end{align}
where $\mathbf{\Phi}^{(l)}\in\mathbb{R}^{D\times D'}$ is the filter matrix of the $l$-th layer.  $\textit{HGNN}_\phi(\cdot)$ can be set as any hypergraph neural networks, such as HGNN~\cite{feng2019hypergraph}, HyperGCN~\cite{yadati2019hypergcn} or HGAT~\cite{yang2021hgat}. In practice, we adopt a one-layer HGNN due to its simplicity and efficiency. As shown in Equation (\ref{eq:hgnn}), $\mathbf{W}_e$ is the weight matrix of hyperedges which we set as the identity matrix here. $\mathbf{D}_v$ and $\mathbf{D}_e$ denote the diagonal matrices of node degrees and hyperedge degrees, respectively. $\mathbf{D}_v(i,i)=\sum_j\widehat{\mathbf{M}}^*_t(i,j)$ and $\mathbf{D}_e(j,j)=\sum_i\widehat{\mathbf{M}}^*_t(i,j)$. Activation function $\sigma(\cdot)$ is set as \textit{PReLU}. The target edge embeddings $\mathbf{Z}_t$ and edge-level context embeddings $\mathbf{Z}_p$ are directly obtained from the $L$-th layer edge embeddings $\mathbf{Z}^{(L)}$: $\mathbf{Z}_t=\mathbf{Z}^{(L)}[M_s+1:M_s+M_{tar},:]$ and $\mathbf{Z}_p=\mathbf{Z}^{(L)}[1:M_{tar},:]$. hypergraph-level context embeddings $\mathbf{Z}_s$ is calculated through average readout:
\begin{equation}
    \mathbf{z}_s=Readout\Bigl(\mathbf{Z}^{(L)}\Bigr)=\frac{1}{M_s}\sum^{M_s}_{i=1}\mathbf{Z}^{(L)}[i,:].\label{eq:hgnn_r}
\end{equation}

\subsection{Anomaly Score Computation}\label{sub:score}
As mentioned in Section \ref{sub:view}, \name discriminates the anomalous nodes/edges from the normal majority based on the disagreement level between the target structures and their surrounding contexts. Here we label the normality as 0 and abnormality as 1. In an ideal scenario, the normal node/edge should fit its contextual neighbors well due to its vast majority, so the predicted score of its discrimination pairs should be close to 0, while the predicted score of anomalous node/edge should be close to 1 due to its irregular and scarce pattern compared to its surroundings. Based on the property above, in the following part, we first introduce a novel discriminator to compute anomaly scores, then elaborate on the loss function.

\xhdr{Cosine Similarity Discriminator} Here we propose a novel discriminator based on cosine similarity. To utilize the anomalous signals from nodes and edges to supervise each other, we swap their context embeddings during the discrimination. Specifically, to compute the target node anomaly score $S_{node}$, we first get $\mathbf{z}_p\in\mathbb{R}^{1\times D_o}$ by average pooling the edge-level context embedding $\mathbf{Z}_p$:
\begin{equation}
    \mathbf{z}_p=\frac{1}{M_{tar}}\sum^{M_{tar}}_{i=1}\mathbf{Z}_p[i,:].
\end{equation}
Then the node anomaly score $S_{node}$ is calculated as follows:
{\small
\begin{equation}
\begin{split}
    S_{node}&=Discriminator\Bigl(\overline{\mathbf{h}}_t,\mathbf{z}_p,\mathbf{z}_s\Bigr)\\
    &=\alpha\Bigl(1-CosSim\bigl(\overline{\mathbf{h}}_t, \mathbf{z}_p\bigr)\Bigr) + \beta\Bigr(1- CosSim\bigl(\overline{\mathbf{h}}_t,\mathbf{z}_s\bigr)\Bigr)\\
    &=(\alpha+\beta)-\alpha CosSim\Bigl(\overline{\mathbf{h}}_t, \mathbf{z}_p\Bigr) - \beta CosSim\Bigl(\overline{\mathbf{h}}_t,\mathbf{z}_s\Bigr), \label{eq:al_be_n}
    \end{split}
\end{equation}
}%
where $1-CosSim\bigl(\overline{\mathbf{h}}_t, \mathbf{z}_p\bigr)\in[0,1]$ represents the patch-level disagreement and $1-CosSim\bigl(\overline{\mathbf{h}}_t,\mathbf{z}_s\bigr)\in[0,1]$ is subgraph-level disagreement. $\alpha\in[0,1]$ and $\beta\in[0,1]$ are two balance factors that balance the importance of patch- and subgraph-level discrimination. $CosSim(\cdot)$ is formed as:
\begin{equation}
    CosSim(\overline{\mathbf{h}}_t, \mathbf{z}_{i})=\frac{\overline{\mathbf{h}}_t\mathbf{z}^T_{i}}{\Vert\overline{\mathbf{h}}_t\Vert\Vert\mathbf{z}_{i}\Vert}, i\in\{p,s\}.
\end{equation}
For clarification, we only discuss the case of a single target node above. In practice, \name is trained in a mini-batch manner with size $\mathcal{B}$. Thus, the objective of node anomaly detection is formalized as:
\begin{equation}
    \mathcal{L}_{node}=\frac{1}{\mathcal{B}}\sum^{\mathcal{B}}_{i=1}S_{i,node},\label{eq:loss_n}
\end{equation}

Similarly, edge anomaly scores can also be computed with the discriminator. Firstly, we get $\overline{\mathbf{H}}_p\in\mathbb{R}^{M_{tar}\times D_o}$ and $\overline{\mathbf{H}}_s\in\mathbb{R}^{M_{tar}\times D_o}$ by concatenate the patch-level context embedding $\overline{\mathbf{h}}_p$ and subgraph-level context embedding $\overline{\mathbf{h}}_s$, respectively.
\begin{align}
    \overline{\mathbf{H}}_p&=\vec{\boldsymbol{1}}_{M_{tar}\times 1}\otimes\overline{\mathbf{h}}_p, \\
    \overline{\mathbf{H}}_s&=\vec{\boldsymbol{1}}_{M_{tar}\times 1}\otimes\overline{\mathbf{h}}_s,
\end{align}
where $\vec{\boldsymbol{1}}_{M_{tar}\times 1}$ is a column vector of all 1s, and $\otimes$ denotes the kronecker product.
Then the anomaly scores $\mathbf{S}_{edge}\in\mathbb{R}^{M_{tar}\times 1}$ is computed as follows:
{\small
\begin{equation}
\begin{split}
    \mathbf{S}_{edge}&=Discriminator\Bigl(\mathbf{Z}_t, \overline{\mathbf{H}}_p, \overline{\mathbf{H}}_s\Bigr)\\
    &=\alpha \Bigl(1-CosSim\bigl(\mathbf{Z}_t, \overline{\mathbf{H}}_p\bigr)\Bigr) +\beta\Bigr( 1- CosSim\bigl(\mathbf{Z}_t,\overline{\mathbf{H}}_s\bigr)\Bigr)\\
    &=(\alpha + \beta)-\alpha CosSim\Bigl(\mathbf{Z}_t, \overline{\mathbf{H}}_p\Bigr) - \beta CosSim\Bigl(\mathbf{Z}_t,\overline{\mathbf{H}}_s\Bigr), \label{eq:al_be_e}
    \end{split}
\end{equation}
}%
Each element of $\mathbf{S}_{edge}$ indicates the anomaly score of a target edge. $\alpha$ and $\beta$ are balance factors that are the same as in Eq. (\ref{eq:al_be_n}). The objective of edge anomaly detection is then formalized as:
\begin{equation}
    \mathcal{L}_{edge}=\frac{1}{\mathcal{B}}\sum^{\mathcal{B}}_{i=1}\frac{1}{M_{i,tar}}\sum^{M_{i,tar}}_{j=1}\mathbf{S}_{i,edge}[j]. \label{eq:loss_e}
\end{equation}

\subsection{Model Training}
Compared with contrastive learning methods~\cite{liu2021anomaly, zheng2021generative}, our model calculates anomaly scores with only positive discrimination pairs without the need for sampling negative pairs, which can largely save the computation cost. The non-contrastive object of our model can be formalized by combining $\mathcal{L}_{node}$ and $\mathcal{L}_{edge}$ as follows:
\begin{equation}
    \mathcal{L}(\theta, \phi)=\frac{1}{2}\bigl(\mathcal{L}_{node}+\mathcal{L}_{edge}\bigr). \label{eq:loss}
\end{equation}
\jie{Due to the distinctive nature of our augmented views, we do not symmetrize the loss as \cite{grill2020bootstrap} and \cite{thakoor2021large} did. The potential performance decline can be compensated by multi-level discrimination pairs introduced in section \ref{sub:score}.} Since our model abandons sampling negative pairs, it may risk an undesired trivial solution where all outputs  ``collapse" to a constant~\cite{chen2021exploring}. Inspired by \cite{grill2020bootstrap} and \cite{thakoor2021large}, we use stop gradient and moving average update to avoid this problem. Specifically, for NAD, the graph view encoder serves as an online encoder whose parameters $\theta$ are updated by following the gradient of the cosine similarity w.r.t. $\theta$, and vice versa for EAD:
\begin{equation}
    \theta\leftarrow optimize(\theta, \gamma, \partial_{\theta}\mathcal{L}(\theta, \phi)),
\end{equation}
where $\gamma$ is the learning rate. Here the updates are computed from the gradients with respect to $\theta$ only, using Adam~\cite{kingma2014adam} optimizer. Then, the hypergraph view encoder serves as a target encoder, whose parameters $\phi$ are updated as an exponential moving average of $\theta$, formed as follows:
\begin{equation}
    \phi\leftarrow\tau\phi+(1-\tau)\theta,
\end{equation}

\begin{algorithm} [t!]
  \caption{Forward propagation of \name }\label{alg:1}
  \KwIn{Attributed graph $\mathcal{G}$; Number of training epochs $T$; Batch size $\mathcal{B}$; Number of evaluation rounds $R$.}
  \KwOut{Node and edge anomaly scoring function $\mathcal{F}_n(\cdot)$ and $\mathcal{F}_e(\cdot)$.} 
  Randomly initialize the trainable parameters $\Theta$ for graph encoder and $\Phi$ for hypergraph encoder;\\
  \tcc{Training stage}
  \For {$t=1,2,\ldots,T$}{
  $\mathcal{B}\leftarrow$ Randomly split $\mathcal{V}$ into batches of size $\mathcal{B}$; \\
  \For{batch $b=(v_1, \ldots, v_\mathcal{B})\in\mathcal{B}$}{
  Extract the subgraph centralized on each node in $b$ as graph view, dual-transform and augment the subgraph as hypergraph view, i.e., $(\mathcal{G}_1, \ldots, \mathcal{G}_\mathcal{B})$ and $(\mathcal{G}^*_1, \ldots, \mathcal{G}^*_\mathcal{B})$; \\
  Calculate node and associated graph view embeddings via Eq. (\ref{eq:g_emb_1}) and (\ref{eq:g_emb_2}); \\
  Calculate edge and associated hypergraph view embeddings via Eq. (\ref{eq:hgnn}) and (\ref{eq:hgnn_r}); \\
  Compute the loss objective via Eq. (\ref{eq:loss_n}), (\ref{eq:loss_e}) and (\ref{eq:loss}); \\
  Backpropagation and update the parameters $\Theta$ with gradients, update the parameters $\Phi$ as the exponential moving average of $\Theta$. \\
  }
  }
  \tcc{Inference Stage}
  \For {$v_i\in \mathcal{V}$}{
  Target edges $e\in \mathcal{E}_i=\mathcal{N}(v_i)$
  \For{$r\in R$}{
  Calculate node anomaly score $S_{node}=\mathcal{F}_{n}(v_i)$ via Eq. (\ref{eq:al_be_n}); \\
  Calculate edge anomaly score $S_{edge}=\mathcal{F}_e(\mathcal{E}_i)$ via Eq. (\ref{eq:al_be_e}); \\
  }
  }  
\end{algorithm}

where $\tau\in(0,1)$ is the decay rate. During the training process, we save the model parameters at the epoch that achieves the best performance. These saved parameters are used for inference. During inference, the anomaly score for each target node and target edge in the graph $\mathcal{G}$ is calculated $R$ times to ensure that the final anomaly scores are statistically stable and reliable.

\subsection{Complexity Analysis}\label{sub:comp}
We conduct a time complexity analysis of the proposed \name, and illustrate its advantage over the existing contrastive learning based methods such as CoLA~\cite{liu2021anomaly} and SL-GAD~\cite{zheng2021generative}. For graph view sampling on target node $v_t$, the time complexity of our h-hop neighbor sampling approach is $\mathcal{O}(\delta N_s)$, where $\delta$ denotes the average node degree in the graph and $N_s$ is the node number of the graph view. Since the dual hypergraph transformation can be completed by simply transposing the incidence matrix of the graph view, its time complexity is $\mathcal{O}(N_s)$. The time complexity of the GNN and HGNN encoders without negative sampling is $\mathcal{O}(M_s)$, while the complexity of the contrastive learning GNN module is $\mathcal{O}(N_s^2)$. We ignore the cost of generating augmentations and scoring anomalies due to their constant complexity. Thus, the total time complexity for \name on a graph with $N$ nodes is $\mathcal{O}(N((\delta+1)N_s+M_s))$, Figure \ref{fig:mem} and Table \ref{tab:time} show an empirical comparison of \name, CoLA and SL-GAD's computational requirements on four benchmark datasets.

\section{Experiment}
In this section, we conduct extensive experiments on six benchmark datasets to evaluate the performance of \name. Specifically, we aim to answer the following questions:
\textbf{RQ1:} Can \name outperform SOTA baselines on both node anomaly detection and edge anomaly detection tasks?
\textbf{RQ2:} Can the detection of anomalous nodes and edges benefit each other and enhance their detection capabilities?
\textbf{RQ3:} How efficient is \name with comparison to SOTA baselines?
\textbf{RQ4:} How do different hyper-parameter values affect the performance of \name?

\subsection{Datasets}
We assess the performance of \name on six benchmark datasets, from a range of domains. Among them, Cora~\cite{sen2008collective}, Pubmed~\cite{sen2008collective}, and ACM~\cite{tang2008arnetminer} are citation networks, which consist of published papers as nodes and citation relationships between them as edges, with each node featuring a bag-of-words representation of the corresponding paper. BlogCatalog~\cite{tang2009relational} and Flickr~\cite{tang2009relational} are social networks, depicting users as nodes and their friendships or followings as edges, with textual content from users’ activities serving as node attributes. DGraph~\cite{huang2022dgraph} is a large-scale financial network, showcasing registered users as nodes, emergency contact relationships as edges, and 17 attributes from users’ personal profiles as node features. Although the original DGraph dataset includes temporal information, we treat it as a static graph for the purpose of this static Graph Anomaly Detection (GAD) study.

\begin{table}[t!]
  \caption{Statistics of the Datasets. NA represents node anomalies and EA represents edge anomalies}
  \begin{tabular}{l|rrrrr}
    \toprule
    \textbf{Datasets} & \textbf{Nodes} & \textbf{Edges} & \textbf{Attributes} & \textbf{NA}& \textbf{EA} \\
    \midrule
    Cora &2,708 &5,429 &1,433 & 150 & 1,232 \\
    Pubmed &19,717 &44,338 &500 & 600 & 7,878\\
    % Citeseer  &2,277 &4,732 &3,703 &6 &0.7362 &0.7175  \\
    ACM  &16,484 & 71,980 & 8,337 & 600& 5,332\\
    BlogCatalog &5,196 &343,486 & 8,189 & 300 & 3,154 \\
    Flickr &7,575 &479,476 &12,047 &450 & 4,729   \\
    DGraph &3,700,550 &4,300,999 &17 &15,509 & 20,312 \\
    \bottomrule
  \end{tabular}
  \label{tab:datasets}
\end{table}
%\footnote{\url{https://pubmed.ncbi.nlm.nih.gov/}} \footnote{\url{https://networkrepository.com/soc-BlogCatalog.php}} \footnote{\url{https://snap.stanford.edu/data/web-flickr.html}} \footnote{\url{https://dgraph.xinye.com/dataset}}
    % \xhdr{Citation Networks} Cora, Pubmed, and ACM are publicly available citation network datasets, where each node represents a published paper, and each edge signifies a citation relationship between two papers. Each node is also associated with a bag-of-words feature of the corresponding paper.
    
    % \xhdr{Social Networks} BlogCatalog and Flickr are popular social network datasets obtained from the BlogCatalog blog-sharing website and the Flickr image-sharing website, respectively. In these datasets, nodes represent users of the websites, while the links represent the following or friendships between users. The textual contents associated with users' activities are considered as node attributes.
    
    % \xhdr{Financial Network} DGraph is a large-scale financial network obtained from a consumer finance industry named ``Finvolution Group", where each node represents a registered user, while an edge between two nodes indicates that a user chooses another user as their emergency contact. Every node is also associated with 17 attributes derived from the basic personal profile of the user. Since we focus on static GAD problem in this work, we disregard the temporal information of the original dataset and treat it as a static graph.

\begin{table*}[!h]
  \setlength\tabcolsep{2.5pt}
  \centering
  \caption{Node anomaly detection performance on six benchmark datasets, the best results on each dataset are in bold. PRE, REC and AUC represent precision, recall and AUC value, respectively.} 
  \label{tab:node_an}
  \begin{tabular}{l|ccc|ccc|ccc|ccc|ccc}
    \toprule
    \textbf{Dataset} & \multicolumn{3}{c|}{\textbf{Cora}} & \multicolumn{3}{c|}{\textbf{Pubmed}} & \multicolumn{3}{c|}{\textbf{ACM}} & \multicolumn{3}{c|}{\textbf{BlogCatalog}} & \multicolumn{3}{c}{\textbf{Flickr}} \\
    \textbf{Metrics} & PRE & REC & AUC & PRE & REC & AUC & PRE & REC & AUC & PRE & REC & AUC & PRE & REC & AUC \\  
    \midrule
    Radar &0.4723 &0.5156 &0.5627 &0.4848 &0.5014 &0.7441 &0.4819 &0.4951 &0.7479 &0.4711 &0.5000 &0.7444 &0.4703 &0.5000 &0.7411 \\
    ANOMALOUS &0.0277 &0.5012 &0.6860 &0.5321 &0.0152 &0.7083 &0.0289 &0.5000 &0.7040 &0.0288 &0.4936 &0.7029 &0.0297 &0.5000 &0.7290  \\
    % \midrule
    DOMINANT &0.5201 &0.5030 &0.7765 &0.0152 &0.5001 &0.8128 &0.4819 &0.4999 &0.8142 &0.5323 &0.5388 &0.6391 &0.5031 &0.5004 &0.7275 \\
    AnomalyDAE &0.5212 &0.5485 &0.7551 &0.7130 &0.5754 &0.7364 &0.7316 &0.6073 &0.7464 &0.6578 &0.5540 &0.7386 &0.5203 &0.5881 &0.7255 \\
    % \midrule
    DGI &0.2408 &0.5273 &0.8100 &0.2315 &0.5210 &0.7153 &0.5228 &0.6365 &0.6154 &0.0289 &0.5000 &0.5781 &0.0297 &0.5014 &0.6189 \\
    CoLA &0.4723 &0.5025 &0.8844 &0.4848 &0.5001 &0.9426 &0.4819 &0.5000 &0.7550 &0.4711 &0.5000 &0.7414 &0.4703 &0.5000 &0.7457 \\
    SL-GAD &0.6195 &0.6845 &0.9016 &0.7470 &0.6027 &0.9218 &0.7213 &0.6319 &0.8146 &0.6809 &0.5641 &0.8054 &0.4937 &0.5021 &0.7664 \\
    \midrule
    \textbf{\name} & \textbf{0.6256}&\textbf{0.7512} &\textbf{0.9116} & \textbf{0.7544} & \textbf{0.7491} & \textbf{0.9561} & \textbf{0.7351} &\textbf{0.7249} &\textbf{0.8285} &\textbf{0.7024} &\textbf{0.7658} &\textbf{0.8145} &\textbf{0.5438} &\textbf{0.6023} &\textbf{0.7821} \\
    \bottomrule
\end{tabular}
\end{table*}

\begin{figure*}[htb]
  \centering
  \subfigure[Cora]{
  \includegraphics[scale=0.31]{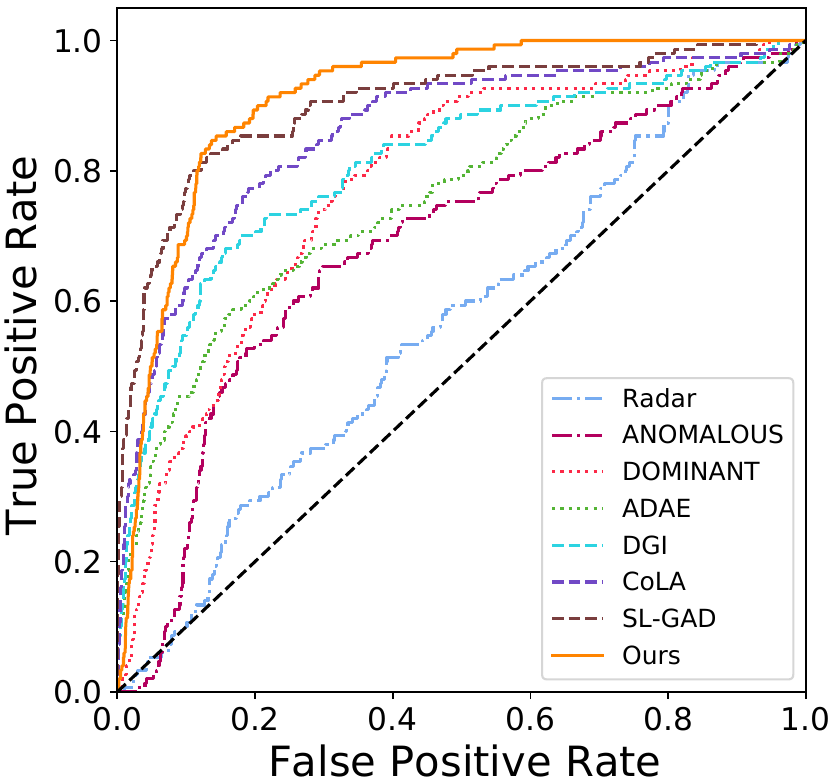}
  }
  \subfigure[Pubmed]{
  \includegraphics[scale=0.31]{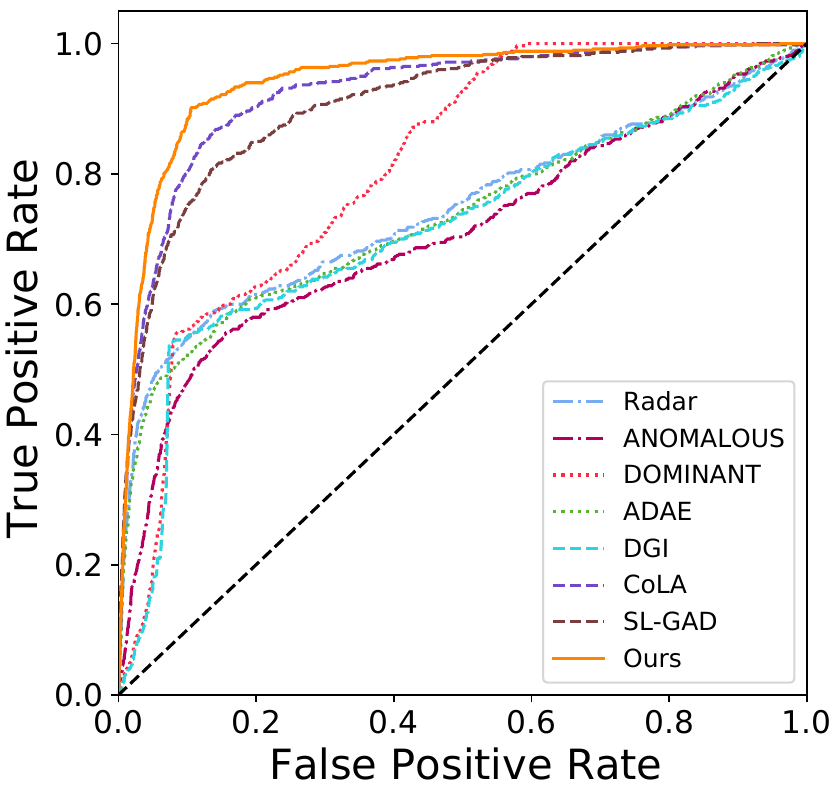}
  }
  \subfigure[ACM]{
  \includegraphics[scale=0.31]{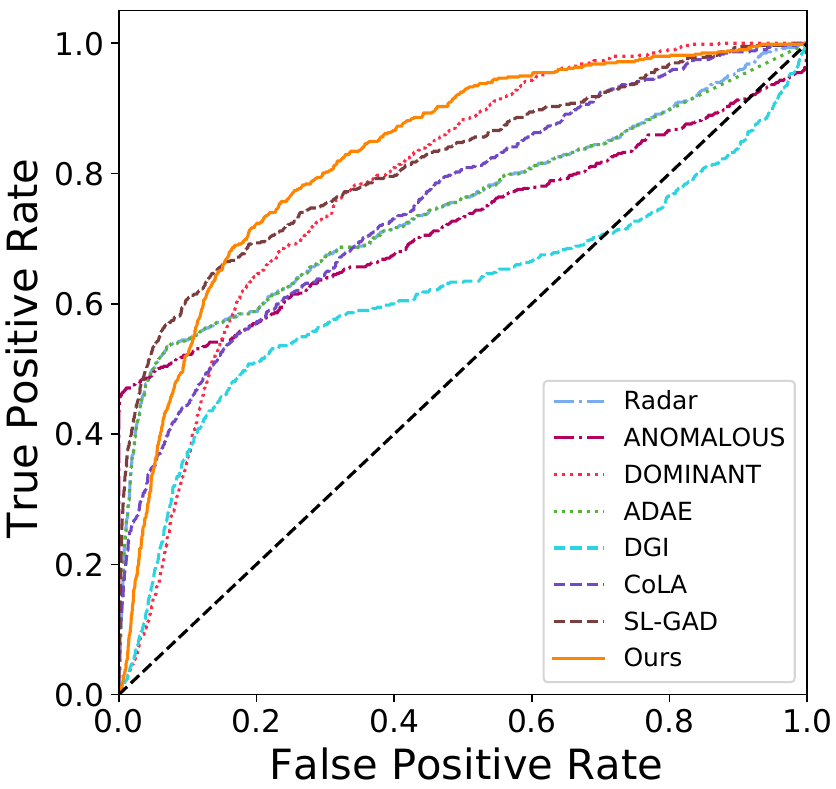}
  }
  \subfigure[BlogCatalog]{
  \includegraphics[scale=0.31]{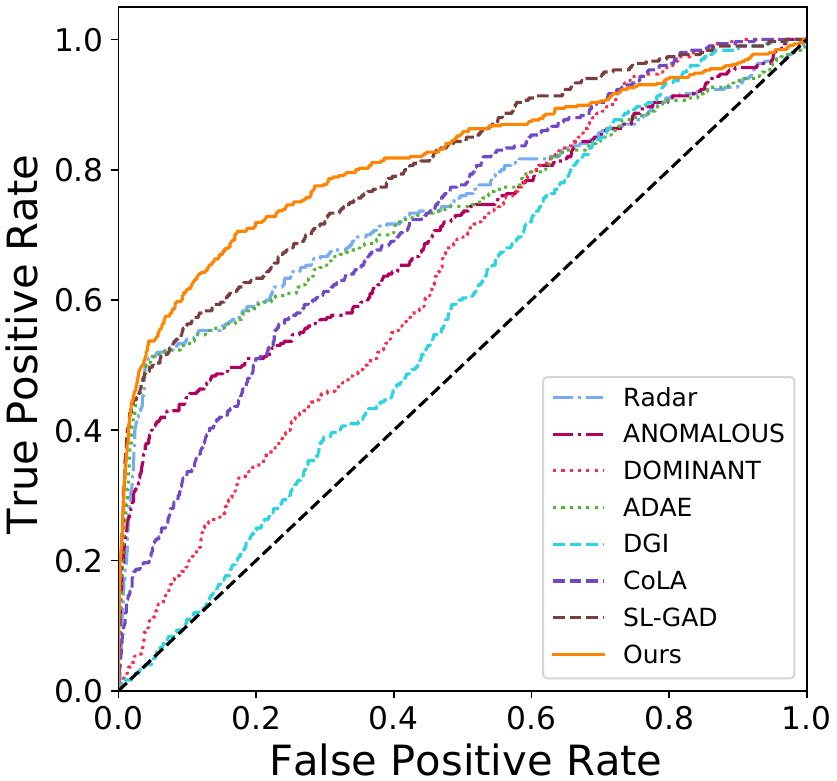}
  }
  \subfigure[Flickr]{
  \includegraphics[scale=0.31]{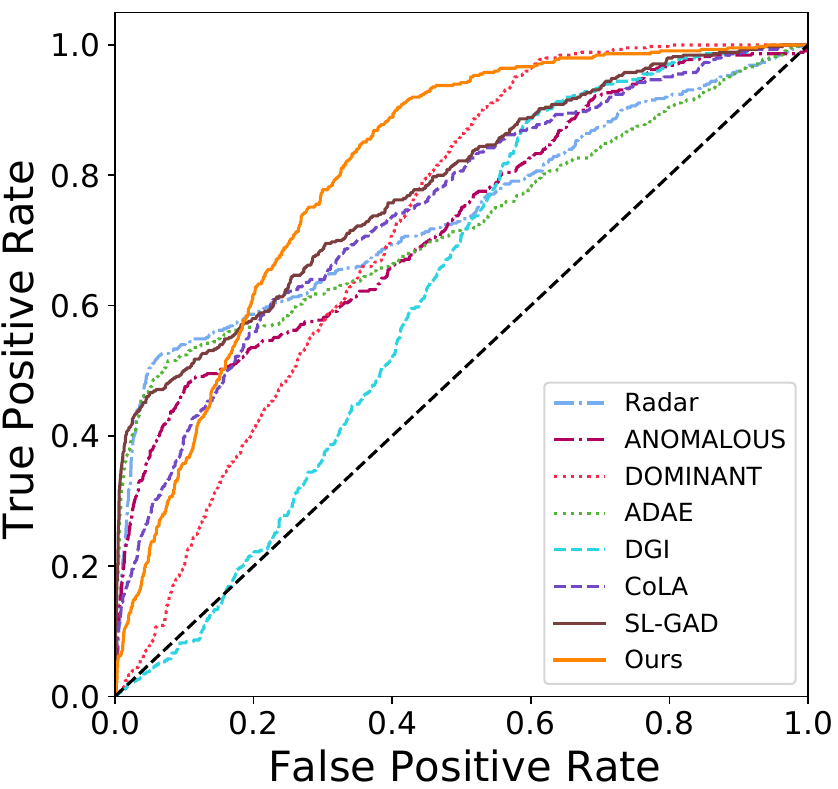}
  }
  \subfigure[DGraph]{
  \includegraphics[scale=0.31]{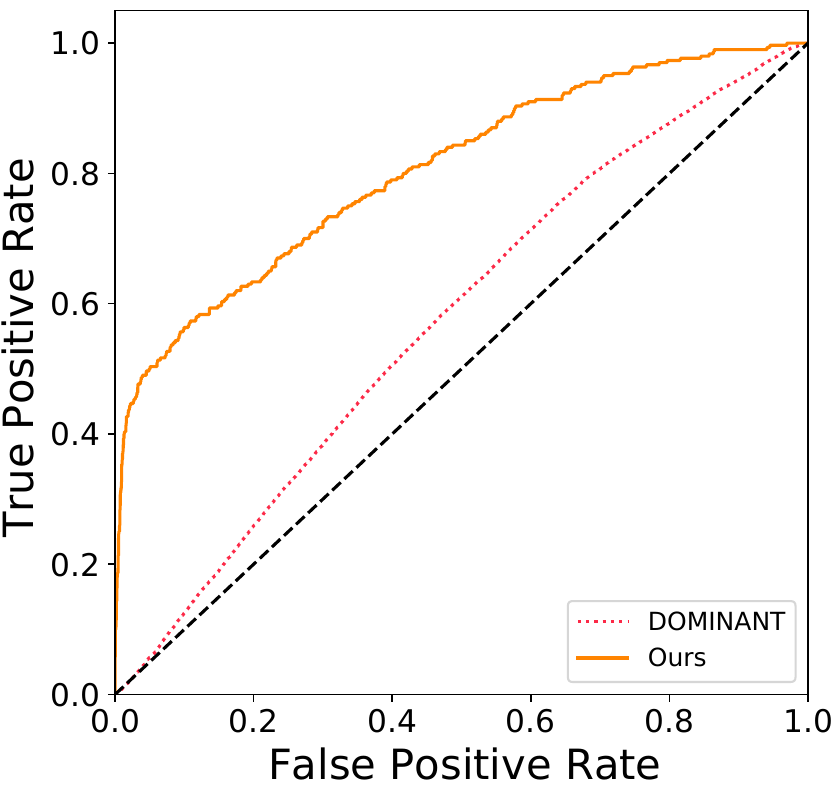}
  }
  \caption{ROC curves on six benchmark datasets for node anomaly detection.} 
  \label{fig:roc_n}
\end{figure*}

The lack of ground-truth labels for both node and edge anomalies is a well-known challenge. Among the six datasets, only DGraph has ground-truth labels for anomalous nodes indicating the fraudulent behaviors of a user. To tackle this problem, we manually inject synthetic anomalies into the original networks for evaluation. Following the anomaly injection strategies used in \cite{ding2019deep} and \cite{liu2021anomaly}, we inject both structural anomalies and attributive anomalies for the other five datasets:
\begin{itemize}
    \item \textbf{Structural anomaly injection.} Following \cite{ding2019deep}, we generate structural anomalies by randomly selecting $n_p$ nodes from node set $\mathcal{V}$ and connecting them to form fully connected cliques. The selected $n_p$ nodes are labeled as structural anomaly nodes, and the newly added edges between them are labeled as structural anomaly edges. This process is repeated $q$ times to generate $q$ cliques. Following \cite{ding2019deep}, we fix $n_p$ as 15 and set $q$ to 5, 200, 20, 10, and 15 for Cora, Pubmed, ACM, BlogCatalog, and Flickr, respectively.
    \item \textbf{Attributive anomaly injection.} Following \cite{liu2021anomaly}, attributive anomalies are created by randomly selecting $n_p\times q$ nodes from $\mathcal{V}$. For each chosen node $v_i$, we sample an additional $2k$ nodes to form two candidate sets: $\mathcal{V}^{n}_i=\{v_1,\ldots,v_k\}$ and $\mathcal{V}^{e}_i=\{v_{k+1},\ldots,v_{2k}\}$. We establish attributive anomaly edges between $v_i$ and $s$ nodes from $\mathcal{V}^{e}_i$ with the largest attribute distances from $v_i$. Then we replace the feature vector of $v_i$ with the node feature from $\mathcal{V}^{n}_i$ that has the largest attribute distance to $v_i$. Following \cite{liu2021anomaly}, we set $s$=2 and $k$=50 for all the datasets.
\end{itemize}

Since DGraph already has realistic labels for anomalous nodes, we only apply attributive anomaly injection with $s$=2 to inject anomalous edges for it. The overall numbers of node and edge anomalies are provided in Table \ref{tab:datasets}.

\subsection{Baselines}
Since \name is a self-supervised GAD model for both node and edge anomaly detection tasks, we compare \name with three types of baselines: (1) Node anomaly detection methods (Radar~\cite{li2017radar}, ANOMALOUS~\cite{peng2018anomalous}, DOMINANT~\cite{ding2019deep}, AnomalyDAE~\cite{fan2020anomalydae}, CoLA~\cite{liu2021anomaly} and SL-GAD~\cite{zheng2021generative}), (2) Edge anomaly detection methods (UGED~\cite{ouyang2020unified} and AANE~\cite{duan2020aane}), and (3) Self-supervised graph representation learning methods (DGI~\cite{velivckovic2018deep} and GAE~\cite{kipf2016variational}). For type (3) methods, since they are not originally designed for anomaly detection, we employ scoring functions from CoLA and UGED to compute anomaly scores for DGI and GAE, repectively. Details of these methods are introduced as follows:
\begin{itemize}[leftmargin=*]
    \item[] \textbf{Node anomaly detection methods:}
\end{itemize}
\begin{itemize}
    \item \textbf{Radar}~\cite{li2017radar} is an unsupervised shallow method that detects node anomalies by analyzing the residual errors in attribute reconstruction.
    \item \textbf{ANOMALOUS}~\cite{peng2018anomalous} is an unsupervised shallow method that detects node anomalies through CUR decomposition and residual analysis.
    \item \textbf{DOMINANT}~\cite{ding2019deep} is a deep graph autoencoder-based unsupervised method that detects node anomalies by assessing the reconstruction errors of individual nodes.
    \item \textbf{AnomalyDAE}~\cite{fan2020anomalydae} is also an autoencoder-based
    unsupervised method. The encoder is graph attention networks~\cite{velivckovic2017graph} while the decoder reconstructs the node attributes. Each node is ranked for anomalies according to its corresponding reconstruction loss.
    \item \textbf{CoLA}~\cite{liu2021anomaly} is a contrastive unsupervised learning-based anomaly detection method that captures node anomaly patterns by measuring the agreement between each node and its contextual subgraph using a GNN-based encoder.
    \item \textbf{SL-GAD}~\cite{zheng2021generative} is a self-supervised anomaly detection method that combines both attribute reconstruction and contrastive learning for detecting node anomalies.
    \end{itemize}
    \begin{itemize}[leftmargin=*]
    \item[] \textbf{Edge anomaly detection methods:}
    \end{itemize}
    \begin{itemize}
    \item \textbf{UGED}~\cite{ouyang2020unified} is an unsupervised edge anomaly detection model that utilizes an autoencoder and a fully connected network to identify edge abnormality through its appearance probability.
    \item \textbf{AANE}~\cite{duan2020aane} is a GNN-based unsupervised edge anomaly detection method that calculates the abnormality of edges as hyperbolic tangents of connected node embeddings.
    \end{itemize}
    \begin{itemize}[leftmargin=*]
    \item[] \textbf{Self-supervised representation learning methods:}
    \end{itemize}
    \begin{itemize}
    \item \textbf{DGI}~\cite{velivckovic2018deep} is a representative graph contrastive learning method that learns node representation by maximizing agreement between positive node-graph pairs and minimizing agreement between negative pairs. We employ the bilinear discriminator in CoLA to score node anomalies.

\begin{table*}[t!]
  \setlength\tabcolsep{2.5pt}
  \centering
  \caption{Edge anomaly detection performance on six benchmark datasets, the best results on each dataset are in bold. PRE, REC and AUC represent precision, recall and AUC value, respectively.} 
  \label{tab:edge_an}
  \begin{tabular}{l|ccc|ccc|ccc|ccc|ccc}
    \toprule
    \textbf{Dataset} & \multicolumn{3}{c|}{\textbf{Cora}} & \multicolumn{3}{c|}{\textbf{Pubmed}} & \multicolumn{3}{c|}{\textbf{ACM}} & \multicolumn{3}{c|}{\textbf{BlogCatalog}} & \multicolumn{3}{c}{\textbf{Flickr}} \\
    \textbf{Metrics} & PRE & REC & AUC & PRE & REC & AUC & PRE & REC & AUC & PRE & REC & AUC & PRE & REC & AUC \\  
    \midrule
    AANE &0.5166 &0.5779 &0.6234 &0.5234 &0.7225 &0.8162 &0.5191 &0.5729 &0.6076 &0.5203 &0.5284 &0.6119 &0.5236 &0.5447 &0.6598  \\
    UGED &0.5230 &0.6072 &0.6672 &0.5414 &0.6875 &0.7471 &0.5030 &0.5567 &0.6388 &0.5194 &0.5250 &0.5869 &0.5276 &0.5575 &0.6491  \\
    GAE &0.4588 &0.4911 &0.5956 &0.5007 &0.5030 &0.5256 &0.5040 &0.5259 &0.5183 &0.5048 &0.4948 &0.5740 &0.5078 &0.5128 &0.5289  \\
    \midrule
    \textbf{\name} &\textbf{0.6623} &\textbf{0.7756} &\textbf{0.8585} &\textbf{0.7367}  &\textbf{0.8928}  &\textbf{0.9765} &\textbf{0.5270}  &\textbf{0.5932} &\textbf{0.7840} &\textbf{0.5558} &\textbf{0.5554} &\textbf{0.7433} &\textbf{0.5508} &\textbf{0.6106} &\textbf{0.8038} \\
    \bottomrule
\end{tabular}
\end{table*}

\begin{figure*}[t!]
  \centering
  \subfigure[Cora]{
  \includegraphics[scale=0.31]{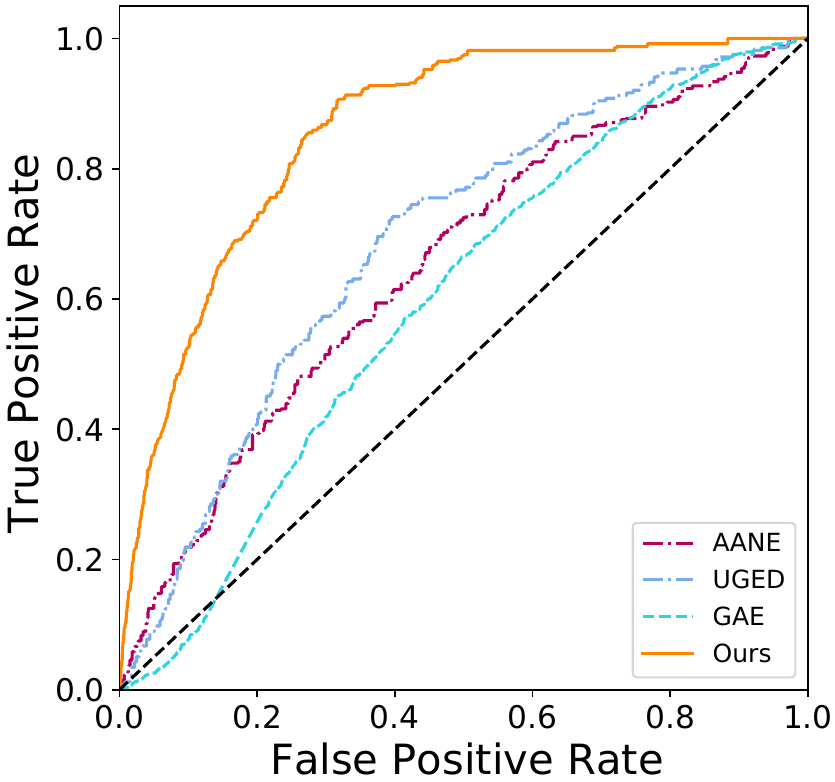}
  }
  \subfigure[Pubmed]{
  \includegraphics[scale=0.31]{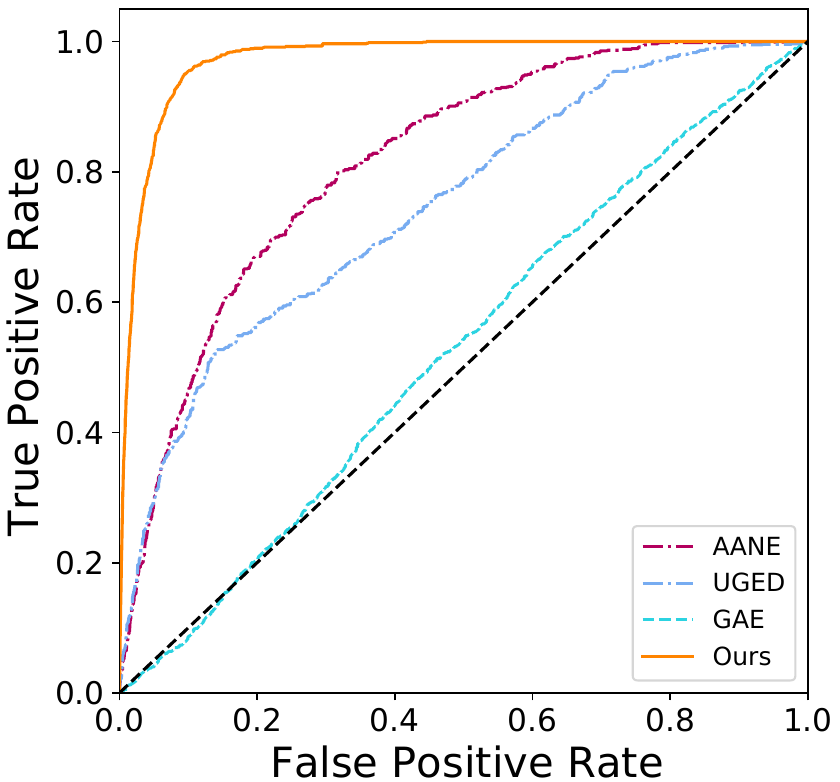}
  }
  \subfigure[ACM]{
  \includegraphics[scale=0.31]{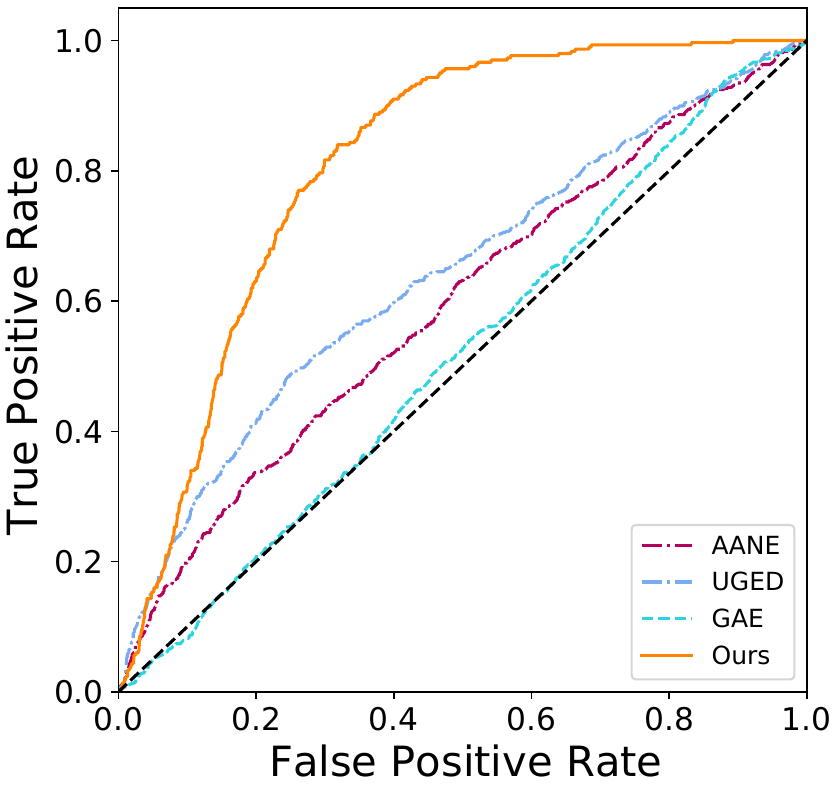}
  }
  \subfigure[BlogCatalog]{
  \includegraphics[scale=0.31]{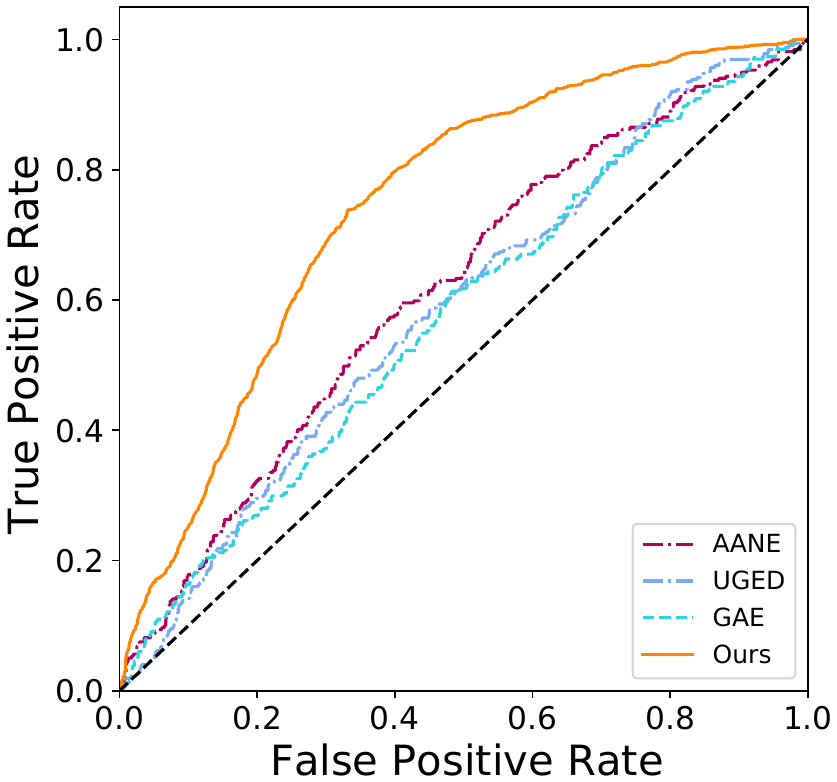}
  }
  \subfigure[Flickr]{
  \includegraphics[scale=0.31]{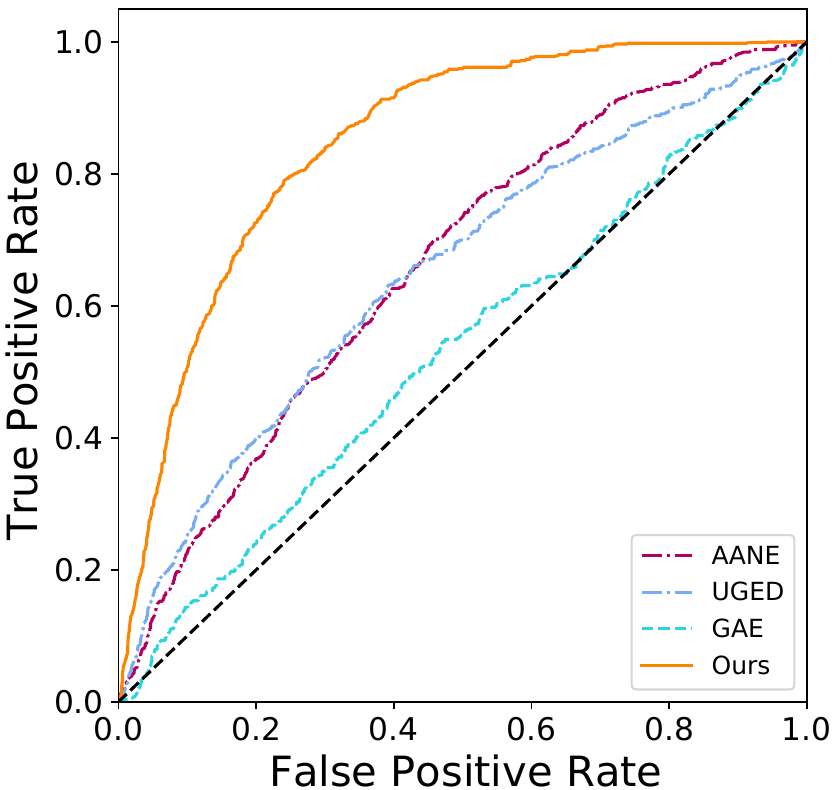}
  }
  \subfigure[DGraph]{
  \includegraphics[scale=0.31]{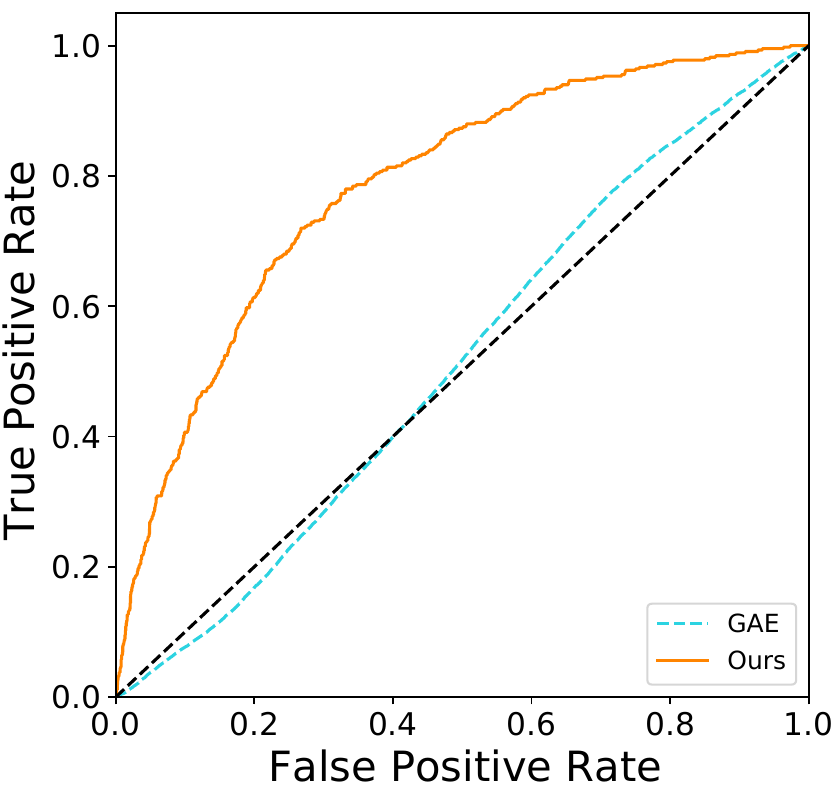}
  }
  \caption{ROC curves on six benchmark datasets for edge anomaly detection.}
  \label{fig:roc_e}
\end{figure*}
    
    \item \textbf{GAE}~\cite{kipf2016variational} learns the latent representation unsupervisedly by minimizing the reconstruction loss between the original graph and the reconstructed graph. Here we further use the strategy from UGED to detect the edge anomalies.
\end{itemize}

\subsection{Parameter Settings} We set the hop size $k$ as 2. The subgraph size $K$ is 40 for BlogCatalog and Flickr and 12 for the rest datasets. Both GNN and HGNN encoders have one layer, with hidden size $D'$ as 128. The hidden size of the predictor is 512. Balance factors $\alpha$ and $\beta$ are selected from $[0.2, 0.4, 0.6, 0.8, 1.0]$. The decay rate $\tau$ is 0.99. The learning rate $\gamma$ is set as 1e-3 for all the datasets. The training epoch is 300 for DGraph and 1000 for the rest datasets. The evaluation round $R$ is 160. For all the baselines, we follow the respective papers to search the suggested hyperparameter spaces to get the best results.

\subsection{RQ1: Node and Edge Anomaly Detection Performance}
To address RQ1, we compare \name with seven baselines for node anomaly detection (NAD) and three baselines for edge anomaly detection (EAD). The precision, recall, and AUC values for NAD are presented in Table \ref{tab:node_an}, while the results for EAD are shown in Table \ref{tab:edge_an}. The ROC curves can be found in Figure \ref{fig:roc_n} (NAD) and Figure \ref{fig:roc_e} (EAD). We only report the ROC curves for DGraph dataset since most baselines run out of memory on it. All the differences between our model and others are statistically significant ($p<0.01$). According to these results, we have the following findings:

\subsubsection{Node Anomaly Detection}
Our proposed \name outperforms all the baselines on all the datasets, showcasing its superiority in node anomaly detection tasks. Compared with the most competitive baselines, \name achieves an average performance gain of 1.48$\%$ and 3.82$\%$ w.r.t. AUC and precision, respectively. Moreover, it demonstrates a substantial improvement of 17.21$\%$ in the average recall. In various real-world applications, such as e-commerce and healthcare, the recall value of anomaly detection is of higher concern.

The shallow methods, Radar and ANOMALOUS, do not achieve satisfactory results due to their limited capacity to handle complex attribute and structure information in graphs. Deep methods exhibit better performance by incorporating strategies like graph reconstruction (e.g., DOMINANT and AnomalyDAE), graph contrasting (e.g., CoLA), or a combination of both (e.g., SL-GAD). However, these methods are limited to a single type of anomaly, i.e., node anomaly. In contrast, \name takes a unified approach by leveraging both node and edge anomaly information simultaneously. This joint consideration enhances the mutual detection of node and edge anomalies and leads to superior performance. 

\subsubsection{Edge Anomaly Detection}
\name also obtains the best performance for edge anomaly detection across all the datasets. The unified detection framework of \name enables the utilization of anomalous node information to enhance the detection of edge anomalies. In comparison to the most competitive baselines, \name demonstrates significant improvements, with an average increase of 15.1$\%$ in precision, 13.86$\%$ in recall, and 22.53$\%$ in AUC for edge anomaly detection. These results highlight the effectiveness of \name in leveraging the correlation between node and edge anomalies.
\subsection{RQ2: Ablation Study}
To further investigate the contribution of each component in \name, we perform an ablation study and present the results in Figure \ref{fig:abl}. We set five variants of \name: \textit{w/o PL}, \textit{w/o SL}, \textit{w/o HGNN}, \textit{w/o GNN} and \name. Among them, \textit{w/o PL} excludes patch-level discrimination by setting $\alpha$=0 and $\beta$=1, \textit{w/o SL} excludes subgraph-level discrimination by setting $\alpha$=1 and $\beta$=0, \textit{w/o HGNN} replaces the HGNN encoder with GNN encoder and focus only on node anomaly detection task, \textit{w/o GNN}  replaces GNN encoder with HGNN encoder and focus only on edge anomaly detection. \name is the full model with all the components available. From Figure \ref{fig:abl}, we have the following observations:
\begin{figure}[t!]
  \centering
  \subfigure[Node Anomaly Detection]{
  \includegraphics[scale=0.28]{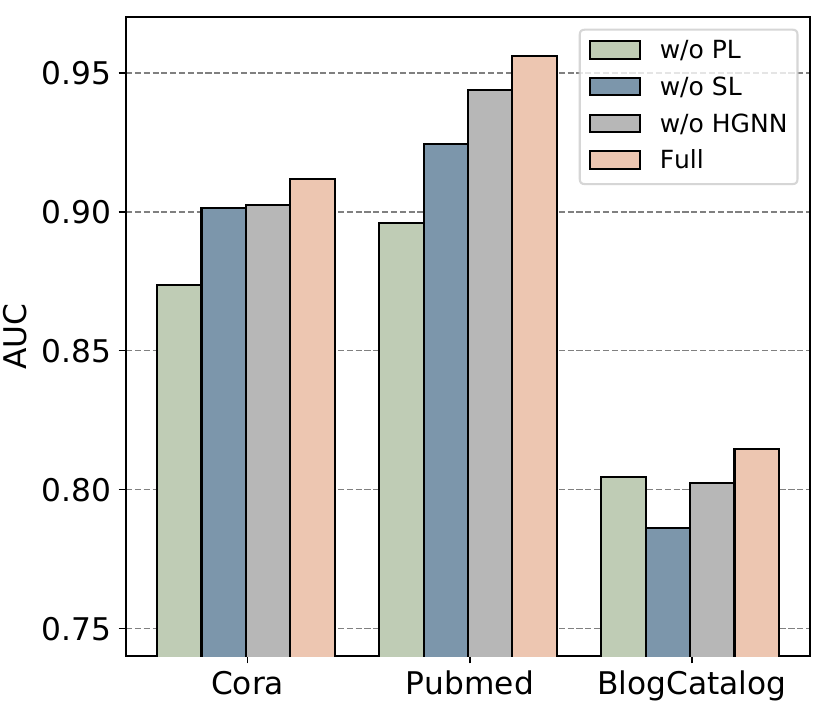}\label{fig:abl_a}
  }\hspace{-2mm}
  \subfigure[Edge Anomaly Detection]{
  \includegraphics[scale=0.28]{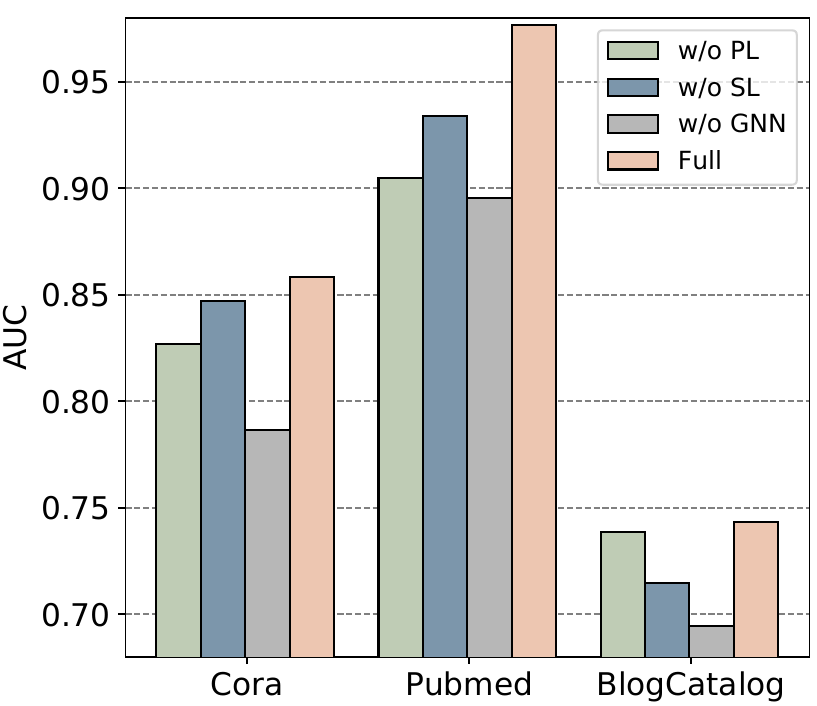}\label{fig:abl_b}
  }
  \caption{Ablation study for NAD and EAD task, respectively.} \label{fig:abl}
\end{figure}

By comparing the results of \textit{w/o PL} and \textit{w/o SL} with \name, as shown in Figure \ref{fig:abl_a} and \ref{fig:abl_b}, we observe a performance decrease on both the NAD and EAD tasks when patch-level or subgraph-level discrimination is removed. This suggests that the combination of patch-level and subgraph-level discrimination is crucial for achieving optimal performance in GAD tasks. By comparing the results of \textit{w/o HGNN} and \textit{w/o GNN} with \name, we observe that the best performance is achieved by the full \name model. This indicates the effectiveness of leveraging node and edge anomaly detection simultaneously. The improved performance of \name confirms that the detection of anomalous nodes and edges can indeed benefit each other for their associations.

% Another observation is that the performance gap between \textit{w/o GNN} and \name is larger compared to that between \textit{w/o HGNN} and \name. This difference can be attributed to the cascading effects of anomalous nodes, which influence the behavior and connectivity of neighboring nodes, thereby causing the presence of anomalous edges. Therefore, the introduction of node anomaly information can greatly enhance the discovery of anomalous edges.

\subsection{RQ3: Model Efficiency Evaluation}
\begin{figure}[b!]
  \centering
  \subfigure[Training Memory]{
  \includegraphics[scale=0.28]{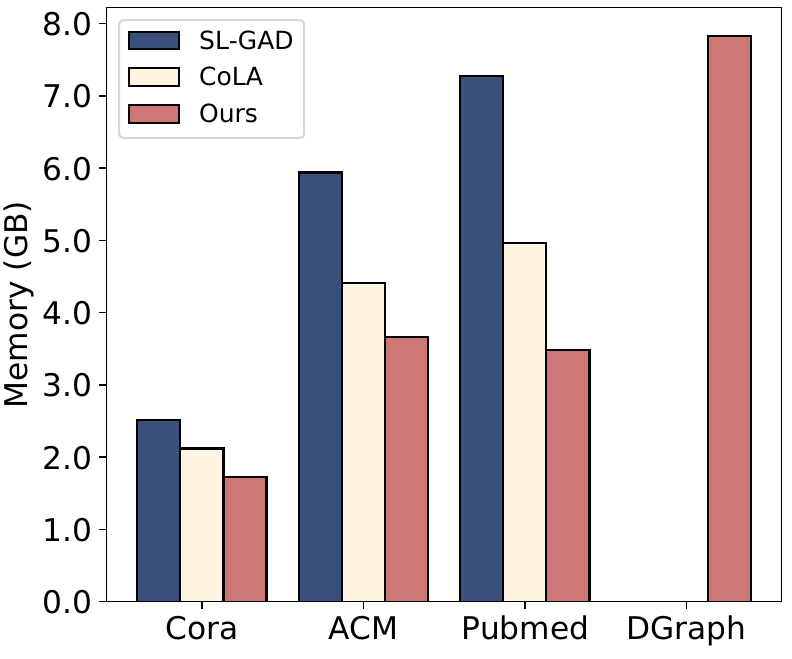}\label{fig:tr_mem}
  }\hspace{-2mm}
  \subfigure[Inference Memory]{
  \includegraphics[scale=0.28]{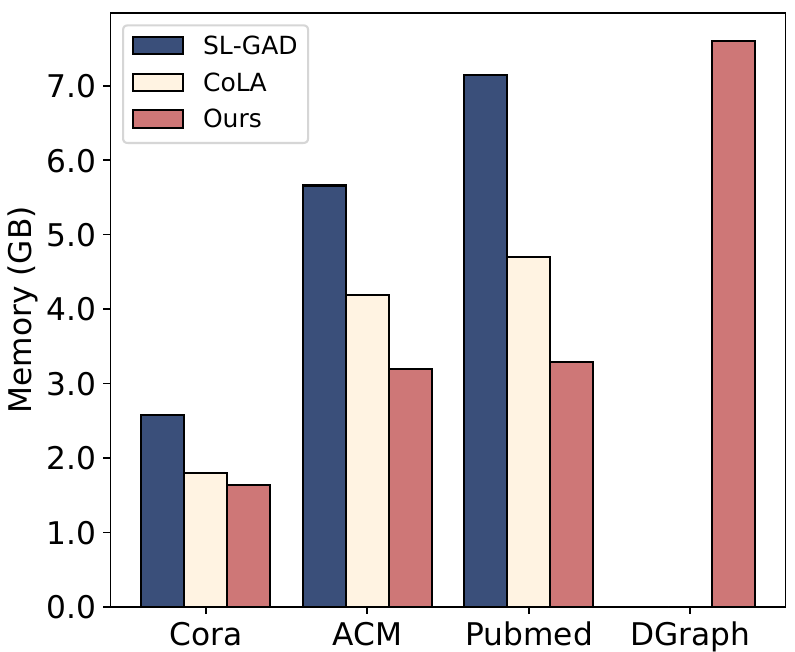}\label{fig:inf_mem}
  }
  \caption{Training and inference memory usage for node anomaly detection.} \label{fig:mem}
\end{figure}

In this subsection, we provide a quantitative analysis of the computational efficiency of our model. Regarding EAD baselines, due to their simple graph auto-encoder based frameworks, they consume much less computational resource than \name. Despite the simplicity, their performance is far from satisfactory, as evident from Figure \ref{fig:roc_e}. Therefore, we mainly compare \name with the two most competitive NAD baselines, CoLA and SL-GAD. To ensure a fair comparison, we use a single-layer encoder with a layer size of 128 for all four models. The training and inference epochs are set to 200 for all. Full batch training is applied for Cora and Pubmed, while the batch size is set to 1,024 and 20,000 for ACM and DGraph, respectively. The memory consumption during training and inference are reported in Figure \ref{fig:mem}, and the training and inference time costs are provided in Table \ref{tab:time}.

% In terms of edge anomaly detection, the baseline UGED, which is a simple graph auto-encoder model, consumes significantly less computational resources than \name. Despite its simplicity, its performance is not satisfactory, as evident from Figure \ref{fig:roc_e}.

As shown in Figure \ref{fig:tr_mem} and \ref{fig:inf_mem}, \name demonstrates lower memory usage during both the training and inference stages compared to the baselines. Additionally, the memory gap between \name and the two baselines is more significant as the dataset size increases. This is attributed to the fact that \name does not require negative pair sampling as SL-GAD and CoLA do, allowing it to easily scale to large graphs. Furthermore, Table \ref{tab:time} illustrates that \name exhibits significantly less training and inference time costs than the baselines. This highlights the superior efficiency of our model, particularly in scenarios that require rapid responses, such as online fraud detection. 

\begin{table}[b!]
  \centering
  \caption{Comparison of computational time on four datasets. OOM indicates running out of memory on a 12GB 2080 GPU. AR indicates acceleration rate.} 
  \begin{tabular}{ll|r|r|r|r}
    \toprule
    \multicolumn{2}{c|}{\textbf{Dataset}} & \multicolumn{1}{c|}{\textbf{Cora}} & \multicolumn{1}{c|}{\textbf{Pubmed}} & \multicolumn{1}{c|}{\textbf{ACM}} & \multicolumn{1}{c}{\textbf{DGraph}} \\
    \midrule
     \multirow{4}{*}{\makecell[c]{Training \\Time}} & CoLA &193.47s &1607.79s & 708.25s &OOM \\
     & SL-GAD &399.32s &3636.15s &1656.73s & OOM \\
     & \textbf{\name} &\textbf{19.97s} &\textbf{85.35s} &\textbf{273.53s} &\textbf{2.72h} \\
     & AR & 9.69 & 18.84 & 2.59 & $-$ \\
     \midrule
     \multirow{4}{*}{\makecell[c]{Inference \\Time}} & CoLA & 182.09s &1518.27s &774.33s &OOM \\
     & SL-GAD &382.76s &3672.24s &1692.15s &OOM \\
     & \textbf{\name} &\textbf{14.37s} &\textbf{58.19s} &\textbf{136.57s} &\textbf{1.25h} \\
     & AR & 12.67 & 20.09 & 5.67 & $-$\\
    \bottomrule
\end{tabular}
\label{tab:time}
\end{table}
\begin{figure*}[t!]
  \centering
  \subfigure[Cora]{
  \includegraphics[scale=0.27]{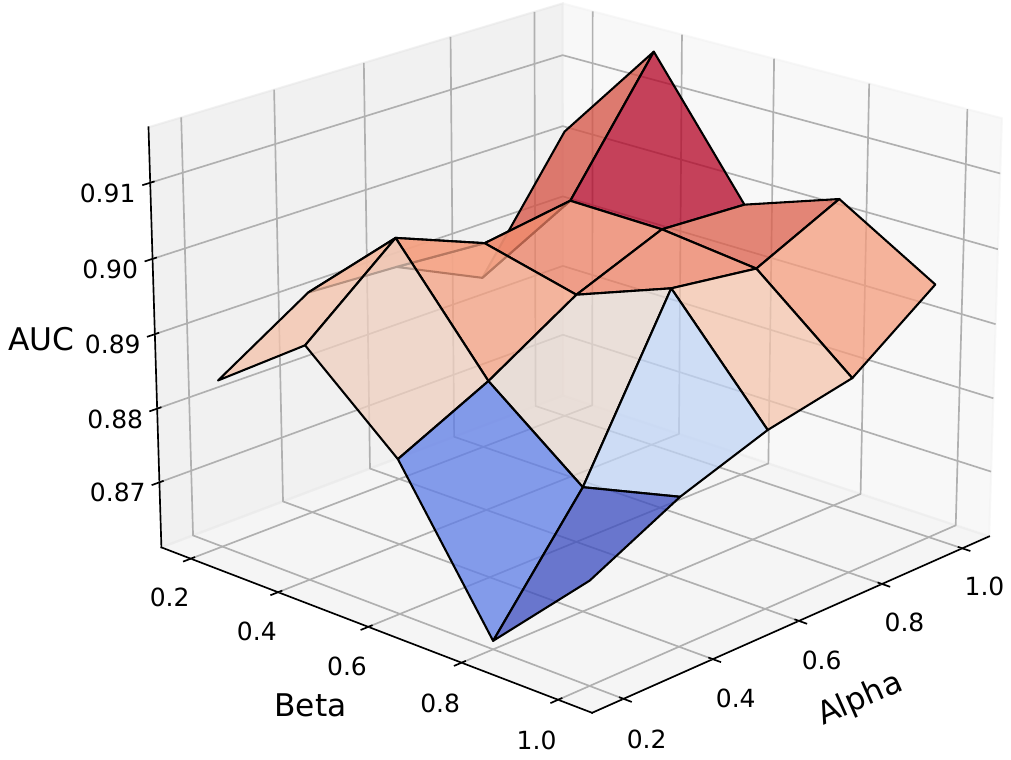}\label{fig:al_be_cora}
  }
  \subfigure[ACM]{
  \includegraphics[scale=0.27]{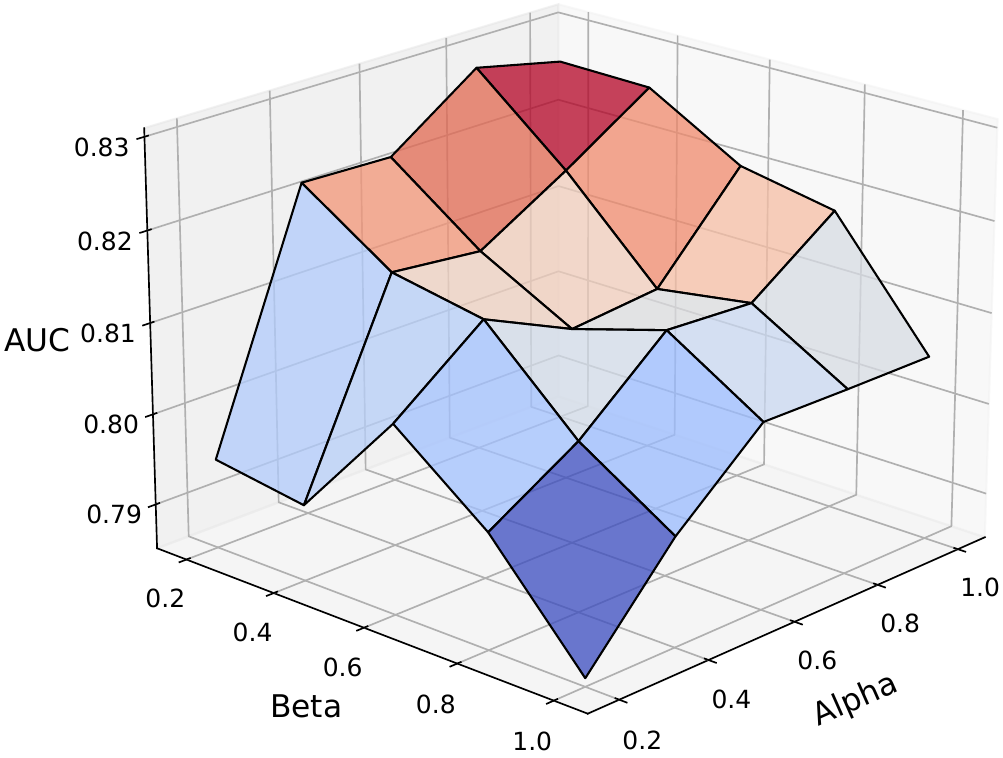}\label{fig:al_be_acm}
  }
  \subfigure[BlogCatalog]{
  \includegraphics[scale=0.27]{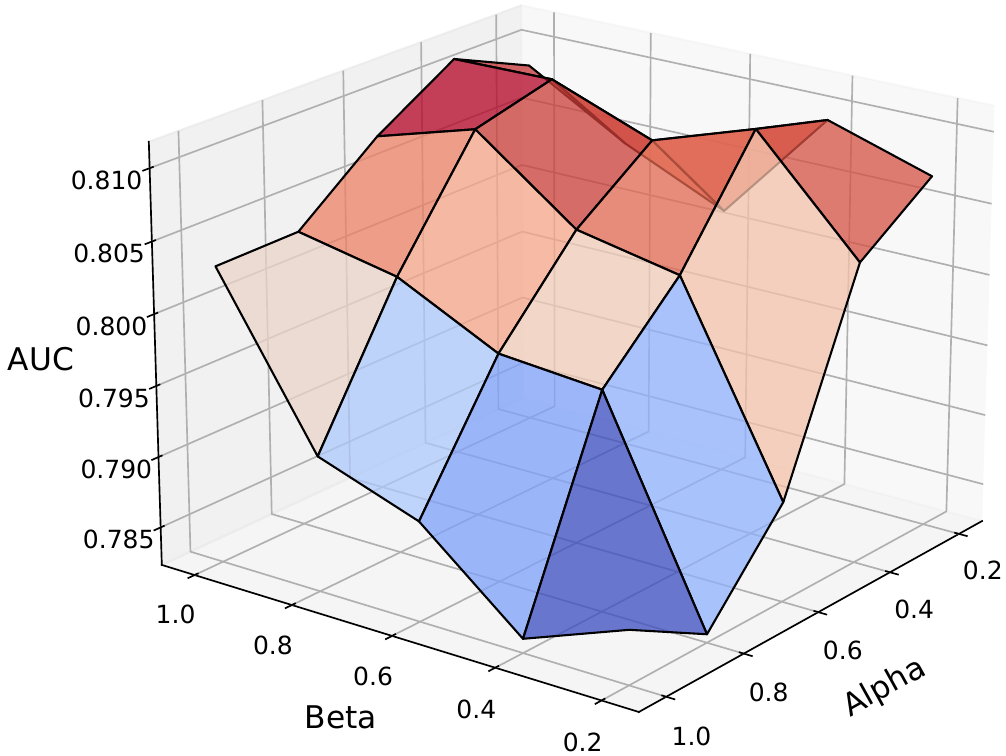}\label{fig:al_be_blog}
  }
  \caption{AUC value of \name on Cora, ACM and BlogCatalog w.r.t. different weights of Alpha and Beta. A warmer color indicates a higher value.} \label{fig:al_be}
\end{figure*}
\begin{figure*}[t!]
  \centering
  \subfigure[Hidden dimension versus AUC]{
  \includegraphics[scale=0.28]{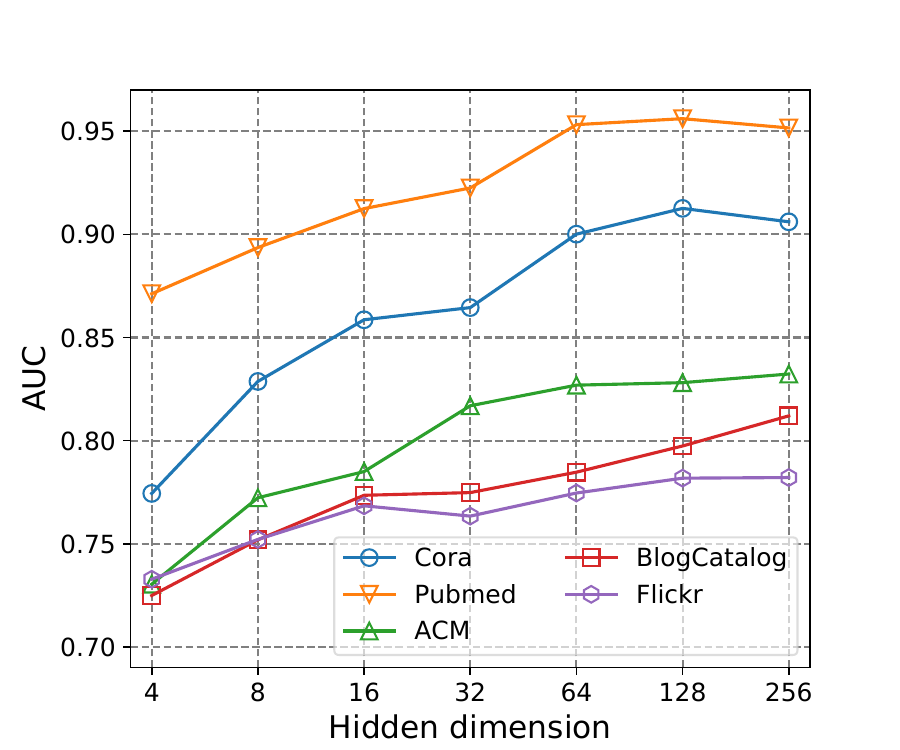}\label{fig:para_hid}
  }\hspace{-2mm}
  \subfigure[Evaluation rounds versus AUC]{
  \includegraphics[scale=0.28]{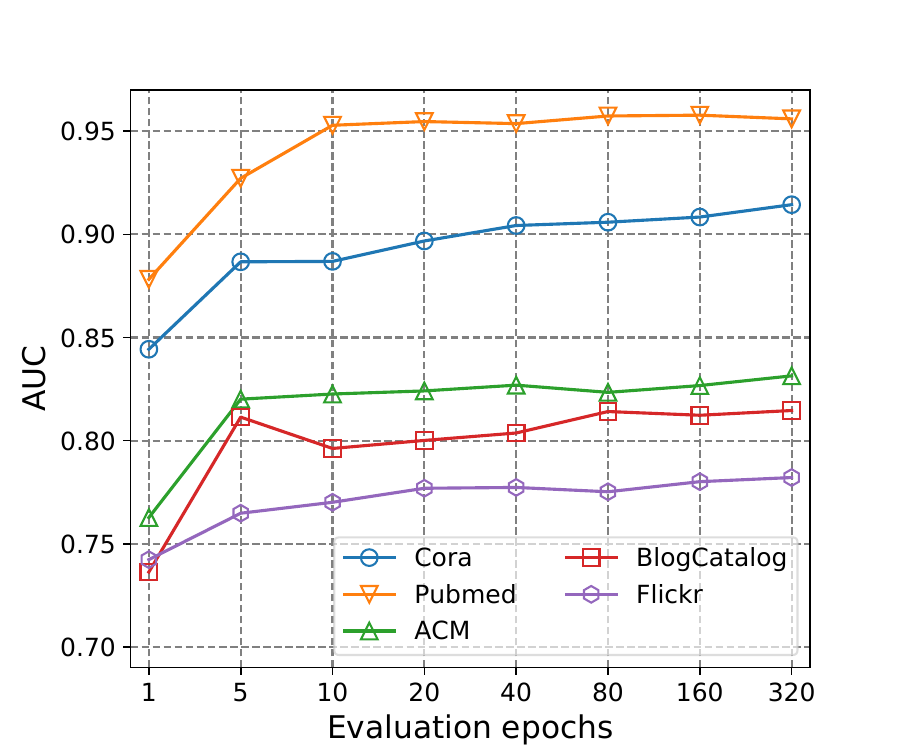}\label{fig:para_evl}
  }\hspace{-2mm}
  \subfigure[Decay rate versus AUC]{
  \includegraphics[scale=0.28]{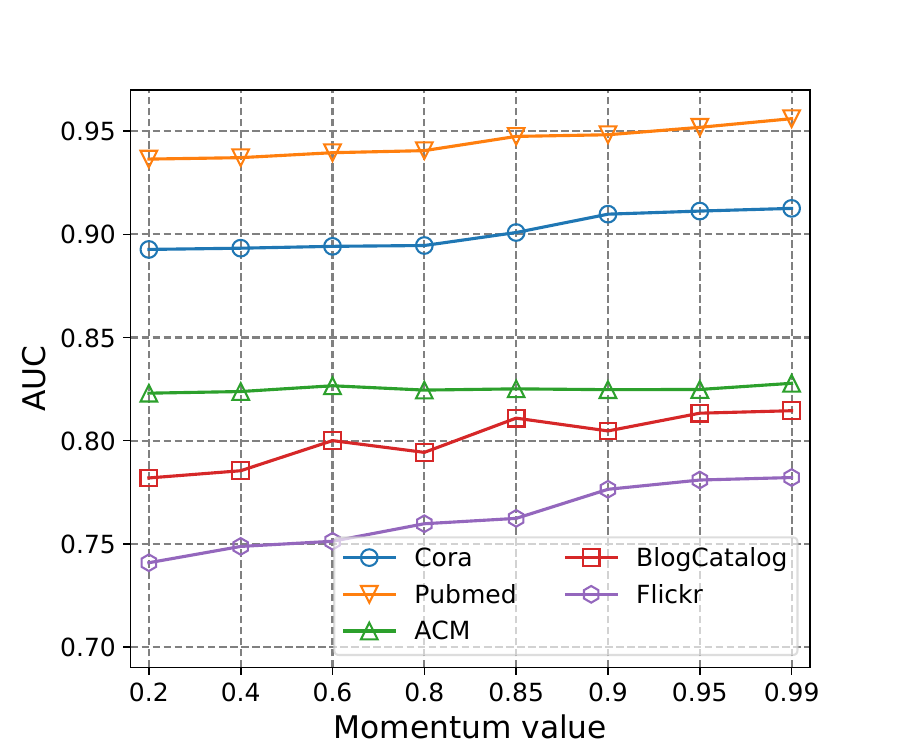}\label{fig:para_mm}
  }\hspace{-2mm}
  \caption{AUC value of \name on Cora, Pubmed, ACM, BlogCatalog, and Flickr w.r.t. hidden dimension $D'$, evaluation rounds $R$ and decay rate $\tau$.} \label{fig:para}
\end{figure*}
\subsection{RQ4: Parameter Sensitivity Analysis}
In this subsection, we conduct a series of experiments to study the impact of various hyper-parameters in \name, including \textit{balance factors}, \textit{hidden dimension}, \textit{evaluation rounds} and \textit{decay rate}.
% the balance factors $\alpha$ and $\beta$ for patch-level and subgraph-level discrimination, the hidden dimension size $D'$, the evaluation epochs \textit{R}, the decay rate $\tau$ for momentum updating and loss weight $\lambda$ which adjusts the importance of two anomaly signals. 
Due to the limitation of space and efficiency, we only show the results for the NAD task on five small-scale datasets, while the EAD task exhibits a similar trend.
\subsubsection{Balance Factor}
In this experiment, we investigate the influence of the balance factors $\alpha$ and $\beta$ in Eq. \ref{eq:al_be_n} and \ref{eq:al_be_e}. We vary both factors within the range of $\{0.2, 0.4, 0.6, 0.8, 1.0\}$ and analyze the corresponding AUC values. The results are presented in Figure \ref{fig:al_be}. According to Figure \ref{fig:al_be_cora} and \ref{fig:al_be_acm}, we observe that the AUC values in citation networks exhibit an increasing trend with higher values of $\alpha$, reaching the peak when $\beta$ is below 0.4. Conversely, in social networks (Figure \ref{fig:al_be_blog}), the AUC values increase with higher values of $\beta$ and achieve the best performance when $\alpha$ is small (below 0.4). This observation suggests that the contribution of patch-level and subgraph-level discrimination is influenced by the characteristics of different datasets. 

\subsubsection{Hidden Dimension}
In this experiment, we investigate the selection of the hidden dimension $D'$ for both HGNN and GNN encoders. We vary the value of $D'$ from 4 to 256 and record the corresponding AUC values. The results, shown in Figure \ref{fig:para_hid}, indicate that the model performance continues to improve as $D'$ increases within the range of [4, 64] for citation networks and [4, 128] for social networks. However, the performance growth becomes marginal as $D'$ further increases. Thus, we set $D'$ as 128 for all the datasets.

\subsubsection{Evaluation Rounds}
To assess the sensitivity of \name to the evaluation Rounds (\textit{R}), we vary \textit{R} from 1 to 320 and examine the corresponding results, as shown in Figure \ref{fig:para_evl}. The figure illustrates that when \textit{R}=1, the detection performance is relatively poor, indicating that too few evaluation rounds are insufficient to capture the abnormality of each node. As \textit{R} increases, most datasets exhibit a significant performance improvement when \textit{R}$\leq 80$. However, further increasing \textit{R} (\textit{R} $\geq$ 160) does not yield substantial performance gains but results in higher computational costs. To strike a balance between performance and efficiency, we set \textit{R} = 160 in our experiments.

\subsubsection{Decay Rate}
In this experiment, we explore the selection of decay rate $\tau$ for the momentum updating procedure. The results are summarized in Figure \ref{fig:para_mm}. In general, we observe that the performance smoothly improves as $\tau$ increases from 0.2 to 0.9 on most datasets (e.g., Cora, Pubmed, ACM and Flickr). Beyond a certain point ($\tau \geq 0.95$), increasing $\tau$ does not significantly affect the performance. Therefore, we set $\tau$ = 0.99 for all the datasets in our experiments.

\section{Conclusion}
In conclusion, our proposed framework BOURNE addresses the limitations of existing GAD methods by introducing a unified approach that detects both node and edge anomalies together. By leveraging the complementary information provided by node and edge anomalies, BOURNE achieves improved performance compared to existing works that treat node and edge anomalies separately. Additionally, our model eliminates the need for negative pair sampling, therefore is highly efficient and scalable for anomaly detection on large graphs. Experimental results on benchmark datasets demonstrate the effectiveness of BOURNE in detecting both node and edge anomalies.

% \section*{Acknowledgment}
% This work is partially supported by The National Key Research and Development Program of China (Grant No. 2020AAA0108504).

\bibliographystyle{IEEEtran}

% Loading bibliography database
\clearpage\bibliography{Unified_GAD}
\vspace{12pt}

\clearpage
\section{Appendix}

\subsection{Further Discussion}
Bootstrapped self-supervised learning is a general self-supervised learning paradigm, obviating the need for negative sampling. While we've built upon foundational frameworks from prior studies~\cite{grill2020bootstrap,thakoor2021large} and adopted a few classic elements (such as stop-gradient and moving average updating), there are key differences between our model and its predecessors and it is crucial to emphasize our contribution and advancements. Prior bootstrapped self-supervised learning models~\cite{grill2020bootstrap,thakoor2021large} are developed for the traditional classification tasks, making them not directly applicable to anomaly detection. To address the challenges, we redesigned some key components to better align with graph anomaly detection.

\begin{figure}[b!]
    \centering
    \includegraphics[scale=0.68]{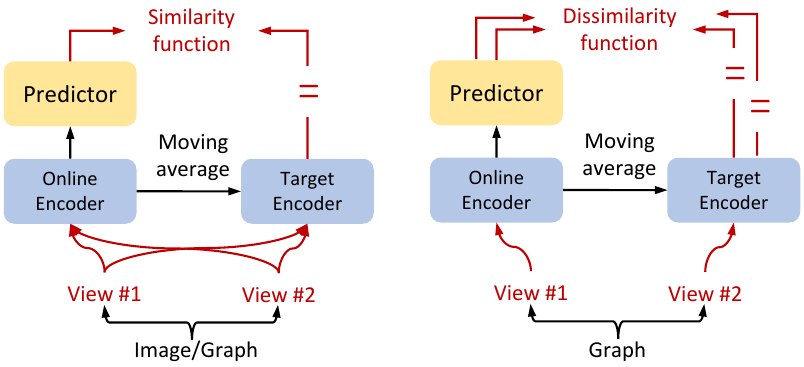}
    \caption{\textbf{Left}: The architectures of BYOL~\cite{grill2020bootstrap} and BGRL~\cite{thakoor2021large}. \textbf{Right}: The architecture of BOURNE. The encoders include all layers that can be shared between both branches. The components in red are those in BOURNE that are different from BYOL and BGRL.}
    \label{fig:dif}
\end{figure}

Specifically, as illustrated in Fig. \ref{fig:dif}, there are three components in BOURNE that differ from the prior architectures: (1) \textbf{Augmentation strategy}. Previous bootstrapped self-supervised models~\cite{thakoor2021large} use edge masking as the augmentation function.
However, this strategy may exclude target edges from the subgraph, introducing extra noises and hindering anomaly detection. To address this problem, we use hyperedge perturbation instead, which only randomly kick out nodes from the hyperedges following an i.i.d. Bernoulli distribution. In this way, the node number in the dual hypergraph remains consistent after the augmentation. We have added the above discussion to section IV-A in the revised paper. (2) \textbf{Symmetrization}. As shown in Fig. \ref{fig:dif}, previous works~\cite{grill2020bootstrap,thakoor2021large} adopt symmetric loss by feeding view \#1 to the target encoder and view \#2 to the online encoder to make additional predictions. This helps boost the accuracy. However, since the augmented views in our model involve both graph and hypergraph, it is not feasible to feed them into the contrary encoders. Thus, we use asymmetric loss but construct both patch-level and subgraph-level pairs for discrimination to compensate for potential performance degradation. We have added the above discussion to section IV-E. (3) \textbf{Similarity function}. As previous works concentrated on image or node classification, they directly calculate the cosine similarity of image/node representations from online and target encoders. However, this approach is unsuitable for anomaly detection, as node or edge abnormalities often hinge on their relationships with neighboring structures~\cite{liu2021anomaly}. To address this, as illustrated in section IV-D, we compute the cosine similarity between target node/edge embeddings and the embeddings of their patch- and subgraph-level contexts. We then use the dissimilarity $(1 - CosSim(\cdot))$ to indicate node or edge abnormality.

\subsection{Hypergraph Perturbation}
We further explicitly outline the motivation for the hypergraph perturbation. The selection of augmentation
strategies can highly affect the performance of self-supervised models. Previous works~\cite{thakoor2021large} commonly use edge masking for graph augmentation. However, this strategy proposed for simple graphs can't apply to hypergraphs. Moreover, this strategy may exclude target edges from the enclosing subgraph, introducing extra noises and hindering anomaly detection. To tackle this problem, we use hyperedge perturbation instead, which only randomly kicks out nodes from the hyperedges following an i.i.d. Bernoulli distribution. This way, the node number within the hypergraph (i.e., edge number in the subgraph) stays constant after augmentation. The hypergraph perturbation module, as an augmentation function, introduces variations in the input graph, helping the model to learn more robust and generalized representations and thus is crucial to model performance. We have done additional experiments to evaluate model performance without hypergraph perturbation. The findings indicate a substantial performance decline without hypergraph perturbation, with AUC on Cora dropping to 0.5524 for node anomaly detection and 0.5609 for edge anomaly detection.

\subsection{Applicability of BOURNE}
We have further studied the performance and applicability of BOURNE by varying the correlation from 0 to 1 in our revision.  Specifically, we first propose a novel metric \textit{anomaly correlation}, $C_{ano}$ to measure the correlations between anomalous nodes and anomalous edges. 
Given an attributed graph $\mathcal{G}$ with anomalous labels $\mathbf{Y}_n$ and $\mathbf{Y}_e$. The anomaly correlation $C_{ano}$ can be defined as the conditional probability $P(e_a\mid v_a)$:
\begin{equation}\small
    P(e_a\mid v_a)=\frac{P(e_a,v_a)}{P(v_a)},
\end{equation}
where $P(e_a,v_a)$ and $P(v_a)$ can be calculated as follows:
{\small
\begin{align}
    P(v_a)&=\frac{\left|\mathcal{V}_a\right|}{\left|\mathcal{V}\right|},\label{eq:yn}\\
    P(e_a,v_a)&=\frac{1}{\left|\mathcal{V}\right|}\sum_{v\in\mathcal{V}}\frac{\left|e\in N(v):y_e=y_v=1\right|}{|N(v)|},\label{eq:ynye}
\end{align}
}%
where $\mathcal{V}_a$ is the anomalous node set. Combining equations (\ref{eq:yn}) and (\ref{eq:ynye}), $C_{ano}\in[0,1]$, we get:
\begin{equation}\small
    C_{ano}=P(e_a\mid v_a)=\frac{1}{\left|\mathcal{V}_a\right|}\sum_{v\in\mathcal{V}_a}\frac{\left|e\in N(v):y_e=y_v=1\right|}{|N(v)|}.
\end{equation}
Graphs exhibiting a strong correlation between node and edge anomalies have a higher value of $C_{ano}$, and vice versa. Then we conduct node and edge anomaly detection w.r.t. varied $C_{ano}\in[0, 1]$. Since the node and edge anomalies are naturally correlated after structural anomaly injection, we only perform attribute anomaly injection for this experiment.
% \begin{table}[!h]
%   \setlength\tabcolsep{2.5pt}
%   % \renewcommand\arraystretch{1.2}
%   \centering
%   \caption{Node anomaly detection performance on six benchmark datasets, the best results on each dataset are in bold.} 
%   \label{tab:node_an}
%   \begin{tabular}{ll|c|c|c|c|c|c}
%     \toprule
%     \multicolumn{2}{c|}{\textbf{$C_{ano}$}} & 1.0 & 0.8 & 0.6 & 0.4 & 0.2 & 0.0 \\
%     \multicolumn{2}{c|}{Metrics} & AUC & AUC & AUC & AUC & AUC & AUC \\
%     \midrule
%     \multirow{2}{*}{NAD} & SL-GAD &0.9130 &0.9078 &0.9172 &0.8899 &0.9025 &0.9029 \\
%     % BOURNE-w/o HGNN & & & & & & \\
%     & BOURNE &0.9393 &0.9215 & 0.9190 & 0.9169 & 0.9067 & 0.9054 \\
%     \midrule
%     \multirow{2}{*}{EAD} & UGED &0.6846 & 0.6756 &0.6651 &0.6735 &0.6787 &0.6690 \\
%     & BOURNE &0.8585 & 0.8503 &0.8160 &0.7844 &0.7542 &0.7463 \\
%     \bottomrule
% \end{tabular}
% \end{table}
\begin{figure}[t!]
  \centering
  \subfigure[$C_{ano}$ versus AUC on NAD]{
  \includegraphics[scale=0.27]{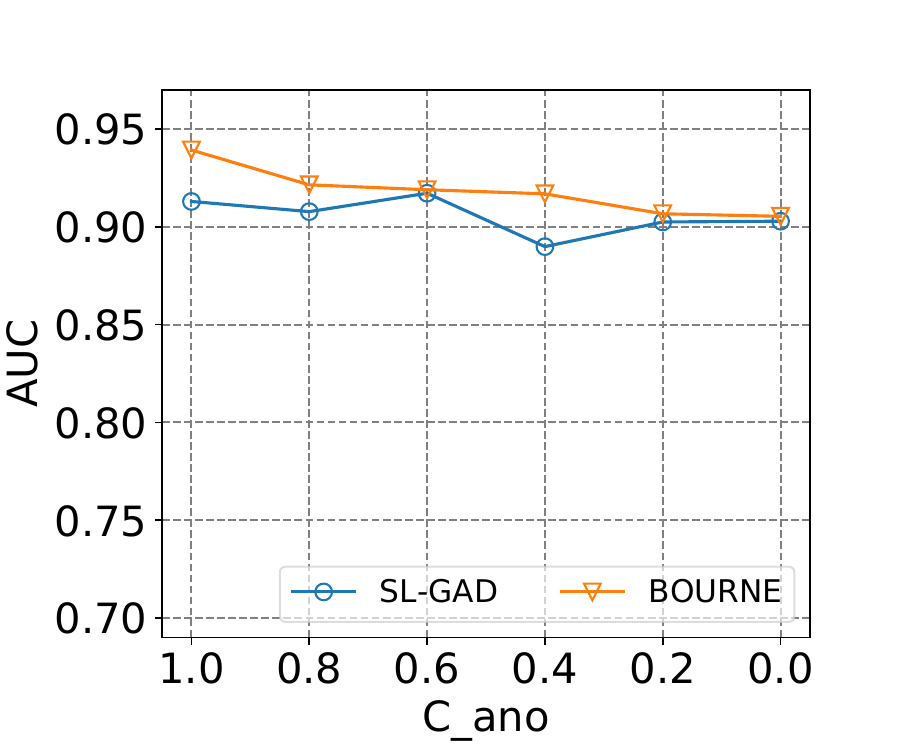}\label{fig:c_node}
  }
  \subfigure[$C_{ano}$ versus AUC on EAD]{
  \includegraphics[scale=0.27]{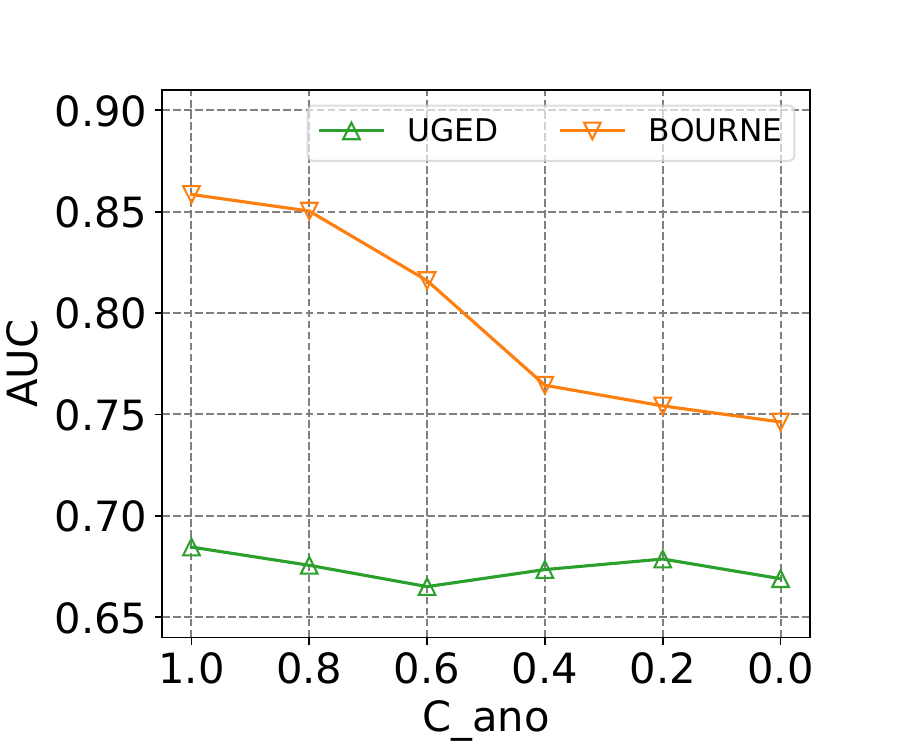}\label{fig:c_edge}
  }
\caption{AUC values of BOURNE and the most competitive baselines (SL-GAD and UGED) w.r.t. varied anomaly correlation $C_{ano}$ on Cora.} \label{fig:c_ano}
\end{figure}

Fig. \ref{fig:c_ano} demonstrates that as $C_{ano}$ decreases, BOURNE's performance in both node and edge anomaly detection gradually reduces, narrowing its performance gaps with the most competitive baselines.  Three observations are made: (1) A strong correlation between node and edge anomalies ($C_{ano}\geq0.6$) enhances the mutual detection of both tasks and boosts the performance. (2) With a loose correlation ($C_{ano}\leq0.2$), the tasks become independent, and the performance enhancement diminishes. However, BOURNE still matches or outperforms the state-of-the-art baselines. (3) Fig. \ref{fig:c_edge} shows that even when $C_{ano}\leq0.2$, BOURNE still significantly surpasses UGED. This is attributed to BOURNE's capability to transform original graph edges into nodes of a dual hypergraph, facilitating explicit edge embedding learning via message-passing. In contrast, UGED only implicitly captures edge information through learned node representations, making BOURNE's edge representations more optimal for anomaly detection. 

\end{document}